\documentclass[twocolumn]{aastex631}

\shorttitle{Merger driven origin of ORCs}
\shortauthors{Dolag et al.}

\graphicspath{{./}{figures/}}

\begin{document}

\title{Insights on the origin of ORCs from cosmological simulations}

\author{Klaus Dolag}
\affiliation{Universit\"ats-Sternwarte, Fakult\"at f\"ur Physik, Ludwig-Maximilians-Universit\"at M\"unchen, Scheinerstr.1, 81679 M\"unchen, Germany}
\affiliation{Max-Planck-Institut f\"ur Astrophysik, Karl-Schwarzschildstr. 1, D-85748, Garching, Germany}

\author{Ludwig M. B\"oss}
\affiliation{Universit\"ats-Sternwarte, Fakult\"at f\"ur Physik, Ludwig-Maximilians-Universit\"at M\"unchen, Scheinerstr.1, 81679 M\"unchen, Germany}
\affiliation{Excellence Cluster ORIGINS, Boltzmannstr. 2, D-85748, Garching, Germany}

\author{B\"arbel S. Koribalski}
\affiliation{Australia Telescope National Facility, CSIRO Astronomy and Space Science, P.O. Box 76, Epping, NSW 1710, Australia}
\affiliation{School of Science, Western Sydney University, Locked Bag 1797, Penrith, NSW 2751, Australia}

\author{Ulrich P. Steinwandel}
\affiliation{Center for Computational Astrophysics, Flatiron Institute, 162 5th Avenue, New York, NY 10010, USA}

\author{Milena Valentini}
\affiliation{Universit\"ats-Sternwarte, Fakult\"at f\"ur Physik, Ludwig-Maximilians-Universit\"at München, Scheinerstr.1, 81679 M\"unchen, Germany}
\affiliation{Excellence Cluster ORIGINS, Boltzmannstr. 2, D-85748, Garching, Germany}

\begin{abstract}

We investigate shock structures driven by merger events in high-resolution simulations that result in a galaxy with a virial mass $M\approx10^{12}$~M$_\odot$. We find that the sizes and morphologies of the internal shocks resemble remarkably well those of the newly-detected class of odd radio circles (ORCs). This would highlight a so-far overlooked mechanism to form radio rings, shells and even more complex structures around elliptical galaxies. Mach numbers of $\mathcal{M} = 2-3$ for such internal shocks are in agreement with the spectral indices of the observed ORCs. We estimate that $\sim$5 percent of galaxies could undergo merger events which occasionally lead to such prominent structures within the galactic halo during their lifetime, explaining the low number of observed ORCs. At the time when the shock structures are matching the physical sizes of the observed ORCs, the central galaxies are typically classified as early-type galaxies, with no ongoing star formation, in agreement with observational findings. Although the energy released by such mergers could potentially power the observed radio luminosity already in Milky-Way-like halos, our predicted luminosity from a simple, direct shock acceleration model is much smaller than the observed one. Considering the estimated number of candidates from our cosmological simulations and the higher observed energies, we suggest that the proposed scenario is more likely for halo masses around $10^{13}$~M$_\odot$ in agreement with the observed stellar masses of the galaxies at the center of ORCs. Such shocks might be detectable with next generation X-ray instruments like the Line Emission Mapper (LEM). 

\end{abstract}

\keywords{Galaxies (573) --- Radio continuum emission (1340) --- Hydrodynamical simulations (767) --- Galaxy mergers (608)}

\section{Introduction} \label{sec:intro}

\begin{figure}[t!]
\centering
\includegraphics[width=0.49\textwidth]{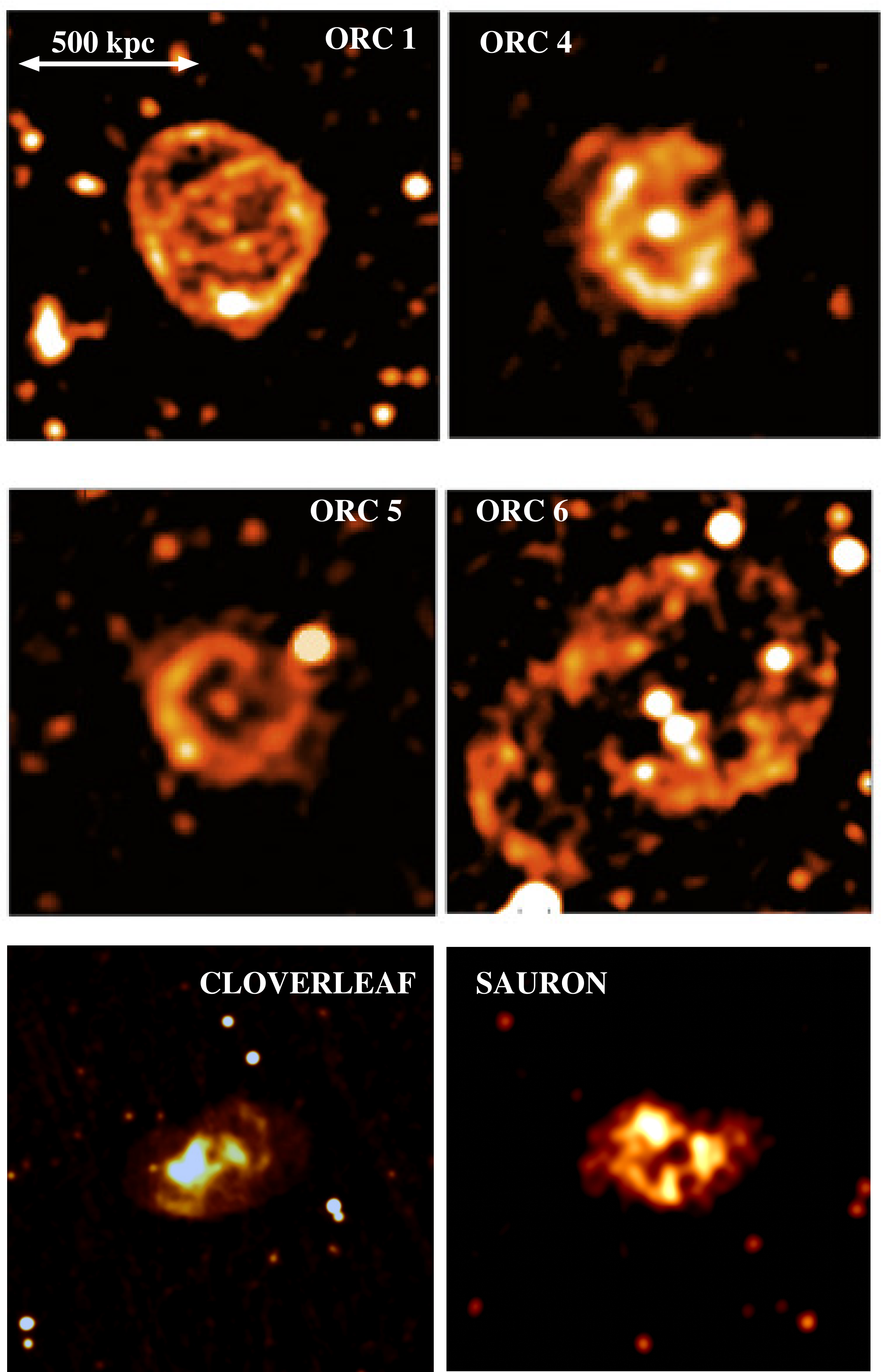}
\caption{ Collection of observed radio continuum emission of odd radio circles (ORCs). The upper row shows a MeerKAT image (left) of ORC~1 \citep{2022MNRAS.513.1300N} and a GMRT image (right) of ORC~4 \citep{2021PASA.38.3N}. The middle row shows an ASKAP image (left) of ORC~5 \citep{2021MNRAS.505L..11K} and a newly discovered ASKAP image (right) of ORC~6, showing a double ring feature (Koribalski et al., in prep.). The lower row shows on the left an ASKAP image of a newly discovered object (CLOVERLEAF, Koribalski et al., in prep.) and on the right a MeerKAT image of an also recently discovered object named SAURON, produced using the data presented in \citet{2022arXiv221102062L}. Each panel has a size of $\approx$ 1~Mpc $\times$ 1~Mpc. \label{fig:orcs_obs}}
\end{figure}

Odd radio circles (ORCs) are a newly discovered class of radio sources, first detected by \citet[][]{2021PASA.38.3N} with the Australian SKA Pathfinder (ASKAP)\footnote{These ASKAP observations are part of the `Evolutionary Map of the Universe' (EMU) Pilot Survey, carried out at a centre frequency of 944 MHz, a bandwidth of 288 MHz and an angular resolution of $\sim$13 arcsec \citep{2021PASA...38...46N}.}. In addition to the three ASKAP-detected ORCs, which consist of one single ORC and one ORC pair, \citet[][]{2021PASA.38.3N} also report a fourth single ORC, detected with the Giant Metrewave Radio Telescope (GMRT) at 325 MHz. The subsequent discovery of a fifth ORC with ASKAP at 944 MHz by \citet{2021MNRAS.505L..11K} made it the third {\em single} ORC around a central elliptical galaxy. Recently, a sixth ORC has been found in ASKAP data (Koribalski et al., in prep.) showing a double ring feature, and another candidate was found by \citet{2022arXiv221102062L} within the MeerKAT Galaxy Cluster Legacy Survey (MGCLS), showing quadrilateral enhanced brightness regions and ear-like features. The currently known single ORCs show large scale, ring-like diffuse radio emission with radii of 150 -- 250 kpc, without any detected counterparts at non-radio wavelength and observed radio powers of several times $10^{23}$ W\,Hz$^{-1}$. The largest linear scale of ORC~6 reaches even $\approx 1$~Mpc in diameter. A collection of ORCs highlighting their different appearances is shown in Figure~\ref{fig:orcs_obs}. Their central host galaxies have redshifts of $z \sim 0.3 - 0.6$ (ORC 1--5, SAURON), but newer candidates are also at much lower redshift like $z=0.125$ (ORC 6) and $z\approx 0.05$ (CLOVERLEAF, Koribalski et al., in prep.). The inferred stellar masses are $M_* \sim 5\times10^{11}$~M$_\odot$, $M_* \sim 2\times10^{11}$~M$_\odot$ and $M_* \sim 10^{11}$~M$_\odot$ for ORC~1, ORC~4 and ORC~5, respectively \citep[][]{2019ApJS..242....8Z}. For ORC~1 \citet{2022MNRAS.513.1300N} report a spectral index of $\alpha$ $\sim 1.4$ (multi telescope) and $\sim 1.6$ (in-band MeerKAT), assuming a flux density $S\propto\nu^{-\alpha}$, while ORCs~4 and 5 have $\alpha \sim 0.92 \pm 0.18$ \citep{2021PASA.38.3N} and $\alpha \sim 0.8 \pm 0.2$ \citep{2021MNRAS.505L..11K}, respectively. For SAURON, \citet{2022arXiv221102062L} report a similar spectral index in the range of $\alpha$ $\sim 1-1.5$.

\begin{table}[t!]
    \centering
    \begin{tabular}{l|l|c|c}
         \hline
         Simulation      & Resolution & $M_{\rm gas}$         & $\epsilon_{\rm gas}$ \\ 
                         &            & [$h^{-1}$~M$_\odot$] & [$h^{-1}$~kpc] \\
         \hline
         COMPASS/asin     & 25000x     & $6.3\times10^3$     & 0.11 \\
         COMPASS/asin     & 250x       & $6.3\times10^5$     & 0.5  \\
         Magneticum/Box4  & uhr        & $7.6\times10^6$     & 1.4  \\
         \hline
    \end{tabular}
    \caption{Summary of the key properties (resolution level, resulting mass of the gas particles and gravitational softening) of the different simulations used.}
    \label{tab:sims}
\end{table}

These different ORC properties have led to numerous speculations on their origin and formation. In the discovery paper by \citet[][]{2021PASA.38.3N} a large number of possible formation scenarios are discussed and mostly considered unlikely. For example, while ORCs resemble supernova remnants, their location at high Galactic latitudes suggests such origin extremely unlikely. The discovery of a third single ORC with a central elliptical galaxy led \citet{2021MNRAS.505L..11K} to favour three possible scenarios for their formation: (1) a relic lobe of a giant radio galaxy seen end-on, (2) a giant blast wave, possibly from a binary supermassive black hole merger, and (3) interactions between a tailed radio galaxy and its environment including neighbouring galaxies. The expected 3D shapes of the radio emission would approximately resemble a cylinder, a sphere, and a ring, respectively. 

Following up on these ideas, \citet{2022MNRAS.513.1300N} investigate whether an expanding synchrotron shell from e.g. a starburst termination shock in the host galaxy can explain the properties of ORC~1,  including its polarisation and spectral index. Interestingly, they find their model broadly consistent with the observations. They also detect little ongoing star formation in the host galaxy, but evidence for a strong starburst event several Gyr ago, and speculate that the ORC’s internal structure may have resulted from galaxy interactions with the starburst wind. There are also investigations by \citet{2022MNRAS.tmpL..74O} which consider if the ejected debris from tidal disruption events, at a rate of approx. one per year over 100 Myr, may generate shocks in post-starburst galaxies \citep[e.g.,][]{2022MNRAS.511.2885B} that resemble ORCs. Although the estimated initial energy of $10^{55}-10^{59}$ erg might be sufficient, this is still small compared to the total energy content of the circum-galactic medium (CGM). It is therefore unlikely that the energy would be sufficient for the shock to reach the observed distance from the host galaxies. Even more exotic suggestions are, for example, relating ORCs to the throats of wormholes \citep{2020EPJC...80..810K}. However, so far none of the discussed mechanisms adequately explains the observed radio structures. A diffuse radio circle discovered in images from the Low Frequency Array (LOFAR) at 144 MHz by \citet{2022RNAAS...6..100O} is most likely the eastern lobe of a giant double-lobe radio galaxy with host WISEA J002034.74+301911.6, an elliptical galaxy at $z = 0.514$ \citep{2012AJ....144..144B}. The system resembles that of the ORC 2+3 pair which appear to be the lobes of a bent radio galaxy with host galaxy WISEA J205848.80--573612.1 \citep[source B in][]{2021PASA.38.3N} at $z \sim 0.3$ \citep{2019ApJS..242....8Z}.

Several projects are underway to find radio sources with rare or unusual morphologies, including the newly discovered class of odd radio circles. For example, using self-organising maps (SOMs, a form of unsupervised machine learning), \citet{2021A&A...645A..89M} scanned the LOFAR Two-metre Sky Survey (LoTSS) images for the highest outlier scores to find sources with rare morphologies, resulting in Wide-Angle Tailed radio galaxies (WATs), Narrow-Angle Tailed radio sources (NATs), relics and halos in galaxy clusters,  but no ORCs were detected. Similarly, \citet{2022arXiv220614677S} searched for the most complex radio sources in the EMU Pilot Survey, only finding two known ORCs. \citet{2022arXiv220813997G}
also apply SOMs to a number of ASKAP fields, including the EMU Pilot Survey, finding the already known ORCs and, in addition, two new ORC candidates. All these attempts confirm that ORCs seem to be a very rare phenomenon.

In this Letter, we use hydrodynamic, non-radiative very-high resolution, cosmological zoom-in simulations of a galaxy halo to investigate the shock structures within the halo and its surroundings as driven by the cosmological assembly of the halo. Similar to what is known from galaxy clusters a short, initial phase is present, where accretion shocks are dominating the virialization of the gas within the halos. Then, around $z \sim 3$ the halo undergoes a transformation and virialization starts to be dominated by internal, merger-driven shocks. Already in early works of idealized galaxy merger simulations it has been pointed out, that merger driven shocks can significantly deploy hot gas of temperatures around $10^6$ and $10^7$ K within the CGM \citep{2004ApJ...607L..87C,2006MNRAS.373.1013C,2009MNRAS.397..190S}, and in fully cosmological simulations contribute significantly to heat the CGM inside galactic halos \citep{2018MNRAS.475.1160H}.
Therefore, as in galaxy clusters, collisions of internal shocks with the accretion shock, so called merger accelerated socks \citep{2020MNRAS.494.4539Z}, push back the accretion shock to several times the virial radius, which can clearly be seen in the simulation of a galaxy halo at $z=6$ as presented in \citet{2020MNRAS.499..597B}.
Subsequent merger shocks are then capable to form complex shock patterns in the outskirts of the halo. We demonstrate that the morphologies of these structures are very similar to the radio emission observed in ORCs, and therefore represent a so-far-unaccounted possibility for their origin. For the first time, we achieve a self consistent model which clearly demonstrates a possible formation mechanism for ORCs and which is able to explain the various observed properties of these systems, including their different morphological appearances as well as their rare frequency of detection in current observations.    

\section{Simulations} \label{sec:sim}

The key parameters of the simulations exploited in this Letter are summarized in Table~\ref{tab:sims}. We mainly used a high-resolution, non-radiative simulation of a galaxy with a final, virial mass of $1.2\times10^{12}$~M$_\odot$ taken from the COMPASS\footnote{\url{http://www.magneticum.org/complements.html##COMPASS}} simulations set, which realizes zoomed initial conditions for various halos from very-massive galaxy clusters down to normal galaxies selected within a 1~Gpc parent simulation \citep[see][]{2011MNRAS.418.2234B}. Here we used the {\it asin} region at {\it 25000x} resolution level\footnote{\href{http://wwwmpa.mpa-garching.mpg.de/HydroSims/Magneticum/Images_Movies/g6802296_shocks_comb.avi}{Splotch movie of {\it asin/25000x}}}, resulting in a gas particle mass of $M_{\rm gas} = 6.3 \times 10^3~h^{-1}$~M$_\odot$ and a softening of $\epsilon = 110~h^{-1}$~pc so that the halo is resolved with $20$ million gas particles within the virial radius at redshift $z=0$. Resolving such shocks at large distances to the center of the halo either needs some shock refinement as used in \citet{2020MNRAS.499..597B}, or, as in our case, employing a very high number of resolution elements to increase the number of particles tracing the low density regions around galactic halos.

The simulation is performed as non-radiative versions using \textsc{OpenGadget3}, which is an advanced version of \textsc{P-Gadget3} \citep{2005MNRAS.364.1105S}, featuring an updated SPH scheme \citep{2016MNRAS.455.2110B}. The high number of particles used allows us to focus on the thermodynamic structures in the outskirts of the halo with very high resolution. Here the simulation contains almost $40$ million gas particles within the accretion shock region and due to the adaptive nature of the simulation, the resolution\footnote{which can be interpreted as the smoothing length of the gas particles, where we use 295 neighbors for our underlying Wendland C6 kernel.} at the accretion radius varies between 10 and 30 $h^{-1}$~kpc and the spatial resolution at the radius of the observed radio ring is typically a factor 3 better. To be able to track the shock structures within the halo, we applied an on-the-fly shock finder \citep{2016MNRAS.458.2080B}. Stellar properties of the galaxy are investigated using a lower spatial resolution simulation, corresponding to a resolution level of {\it 250x}, including galaxy formation physics \citep{2010MNRAS.405.1491M,2015MNRAS.447..178M, 2017MNRAS.470.3167V, 2020MNRAS.491.2779V} as well as selecting galaxies with similar merger history from the {\it Box4/uhr} of the {\it Magneticum Pathfinder} simulation \citep[see e.g.][for a detailed description]{2015ApJ...812...29T}, which corresponds to {\it 25x} in the zoomed simulations. To model synchrotron emission we use \textsc{Crescendo}, a Fokker-Planck solver for spectral cosmic rays in cosmological simulations  \citep{2022arXiv220705087B}, which we applied in a post-processing mode. 

\begin{figure}[t]
\centering
\includegraphics[width=0.49\textwidth]{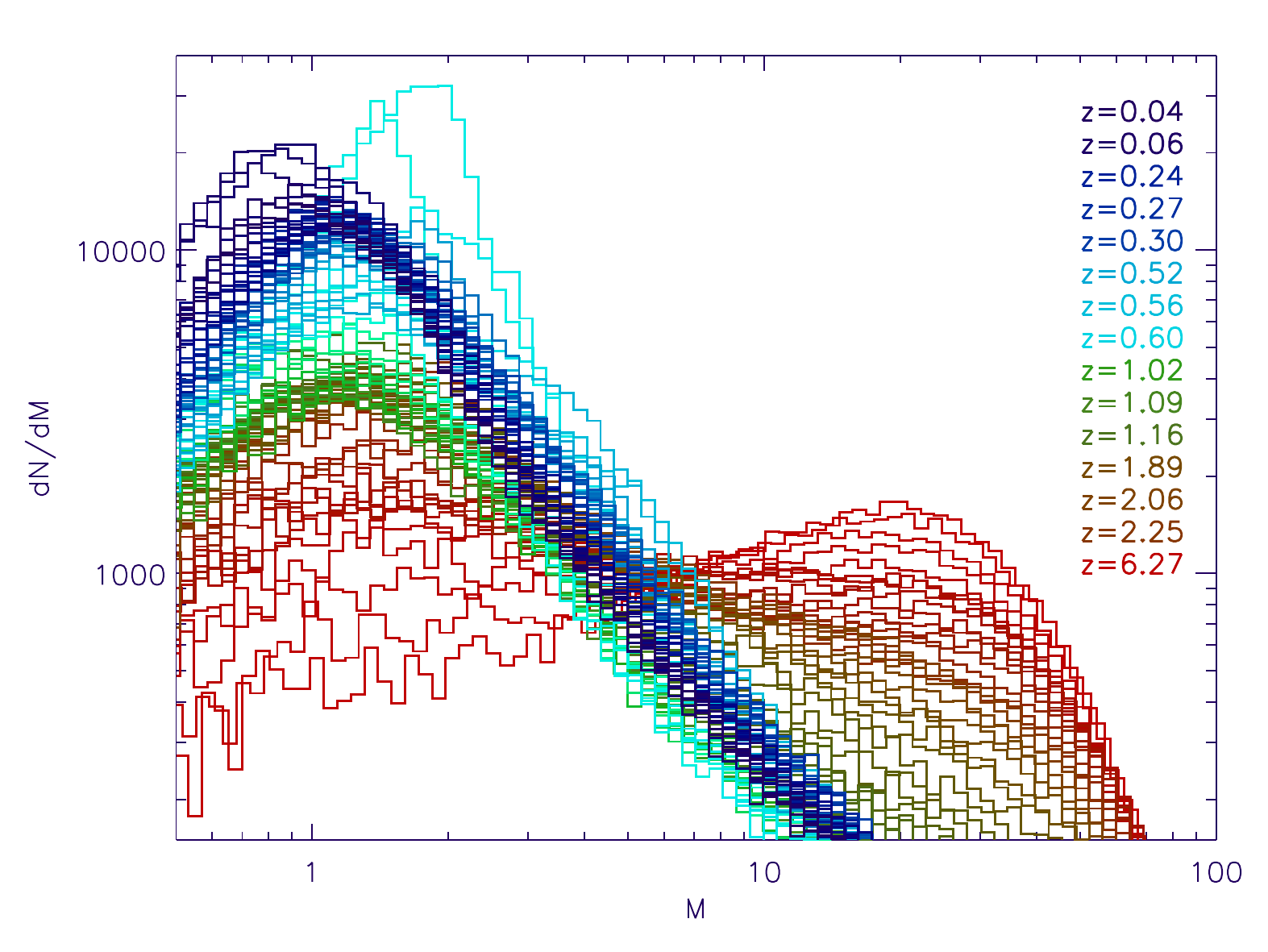}
\caption{We show the evolution of the mass-weighted distribution of the Mach number $\mathcal{M}$ of the detected shocks within a 1 $h^{-1}$cMpc sphere around the galaxy halo. 
The colors represent the redshift, as indicated in the labels. The transition from the accretion shock dominated regime at high redshift to the internal shock dominated regime at low redshift can be clearly seen.   \label{fig:shock_stat}}
\end{figure}

\section{Shocks} \label{sec:shocks}

Similar to galaxy clusters, also the virialization of the galactic halo is driven by accretion shocks only at high redshift. During the main formation of the halo, the heating is dominated by internal shocks, driven by the assembly of the galaxy. Figure~\ref{fig:shock_stat} shows a histogram of the resolution elements with their detected Mach number from the internal, on-the-fly shock finder \citep{2016MNRAS.458.2080B}. At the earliest time shown ($z\approx7$), the distribution clearly peaks at high sonic Mach numbers ($\mathcal{M}>20$), as typical for accretion shocks. Around $z\approx3$ the distribution transforms and internal shocks peaking at $\mathcal{M}\approx2-3$ start to dominate. Below redshift $z\approx2$ there is no longer any sign of a distinguished population of resolution elements tracing the accretion shocks. Our findings here are perfectly aligned with the results for a galactic halo at $z=6$, as presented in \citet{2020MNRAS.499..597B}, both, in the distribution of the Mach number for the accretion shocks with $\mathcal{M}>20$ and pushed back significantly beyond the virial radius already, as well as a peak of internal shocks with Mach numbers $\mathcal{M}\approx 2-3$. Similar to what is found in galaxy clusters \citep{2020MNRAS.494.4539Z}, the collision of internal, runaway shocks with the accretion shock pushes the accretion shock with time far beyond the virial radii. 

\begin{figure}[t]
\centering
\includegraphics[width=0.49\textwidth]{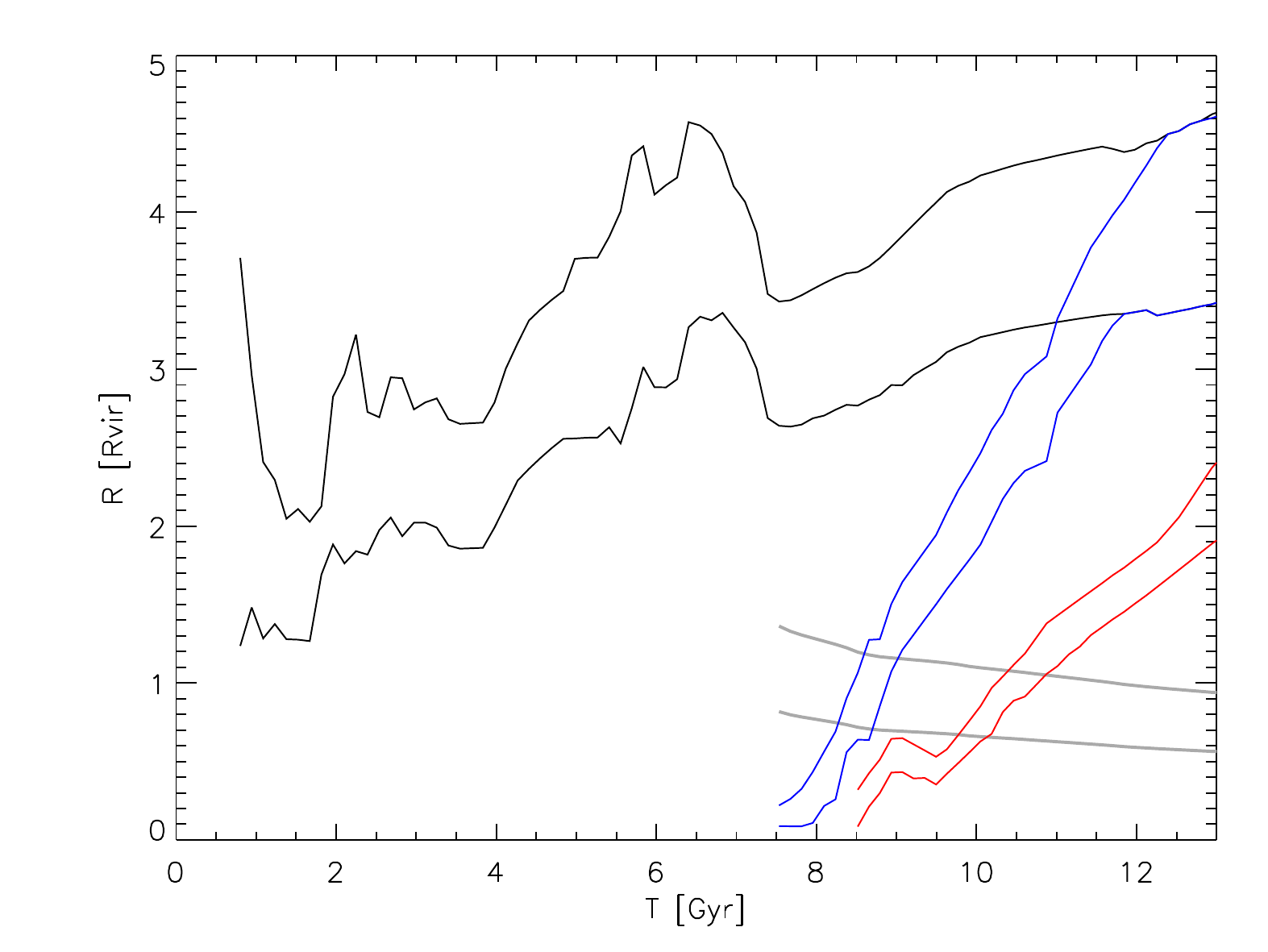}
\caption{We show the evolution of the distance of the different shocks from the halo center in units of the virial radius. The black lines indicate the accretion shock while the blue and the red lines mark the two internal shocks (see text for details). The two gray lines indicate diameters of 300 and 500 kpc, respectively.  \label{fig:shock_radii}}
\end{figure}

In Figure~\ref{fig:shock_radii} the two black lines show the distance of the accretion shock and its evolution with time. As the shape of the accretion shock is quite irregular, ranging to much larger distances towards voids and smaller distances towards in-flowing filaments, the two lines mark a minimum and maximum distance of the detected accretion shock. Already at early times, when the universe was younger than 1~Gyr, the accretion shock in the progenitor system of the galaxy is pushed beyond the virial radius. At $z\approx0.7$ (and shortly after) the formation of the halo leads to a pair of internal shocks (red and blue lines) which are traveling outwards. One of them extends even up to the accretion shock which at late times sits at a distance of $\approx4 R_{\mathrm vir}$. The two lines for each of the shocks again mark the minimum and maximum distance of the shock surface from the center of the halo. The gray lines mark the radial distance of circles with diameter of 300 and 500 kpc, respectively, and already indicate that the size of the observed ORCs can be easily matched by these internal shocks within galaxies halos, even if the larger one roughly correspond to the virial radius of our simulated galaxy. It is interesting to note that these internal shocks also clearly stick out in the overall distribution of shocks in the halo, as can be seen in the histograms presented of Figure~\ref{fig:shock_stat}, where the light blue lines correlate with this time interval.

\begin{figure}[ht!]
\hspace{0.25cm}
\includegraphics[width=0.45\textwidth]{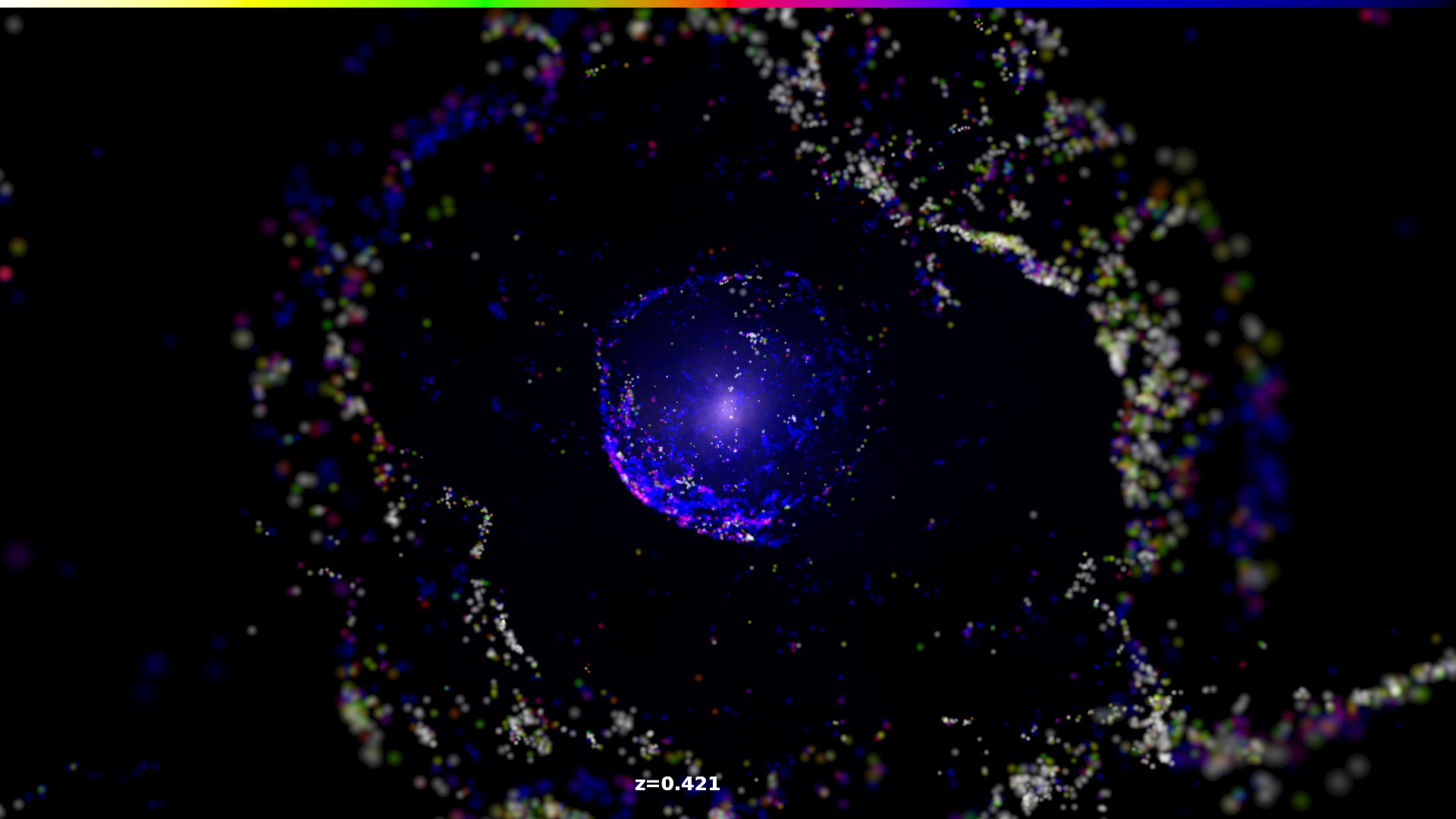}\\
\includegraphics[width=0.48\textwidth]{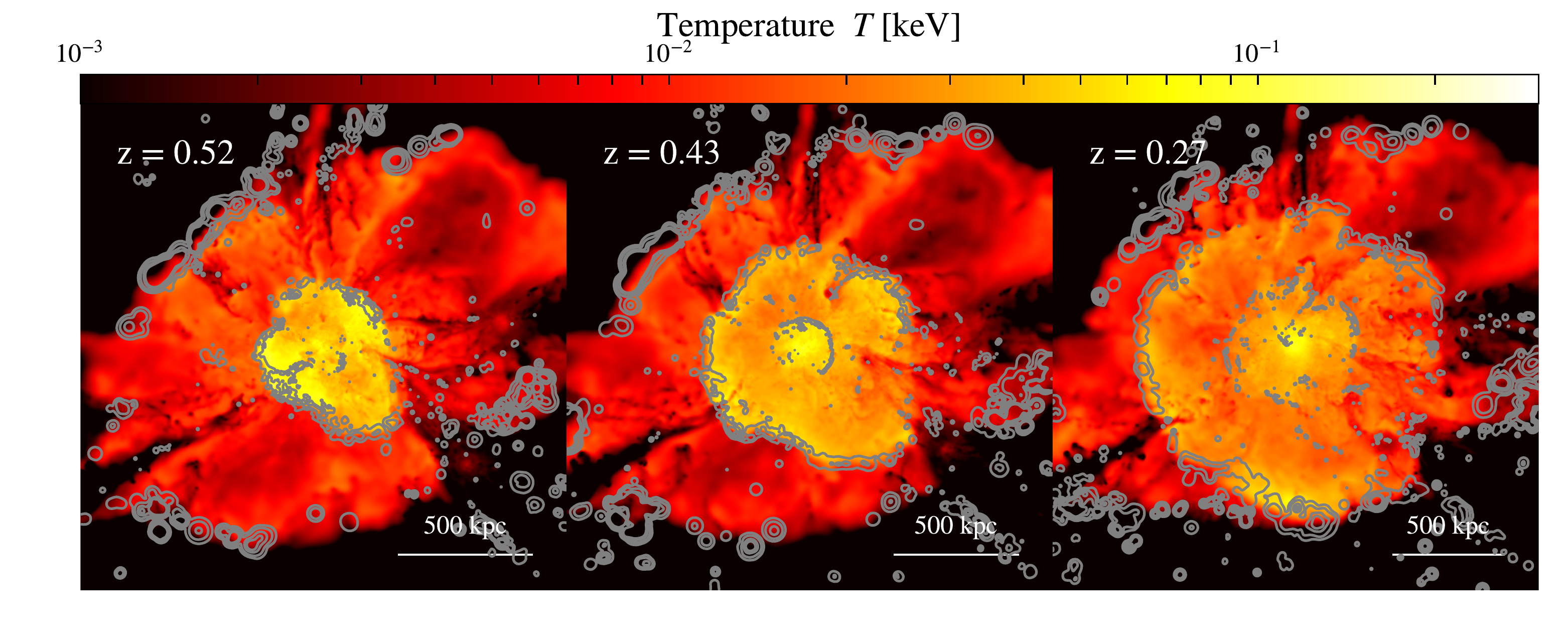}
\caption{Upper panel: Ray tracing movie$^3$ of a 1.3 $h^{-1}$cMpc region using Splotch \citep{2008NJPh...10l5006D}. Shown is a composition of the hot gas (mostly visible in the central part, color coded by temperature) with  the individual, shocked resolution elements color-coded by their mach number $\mathcal{M}$ (where blue, red, yellow and white correspond to $\mathcal{M}\approx 2,3,4$ and $>5$, respectively). Lower panel:
We show maps of a 200 $h^{-1}$ckpc thin slice of the IGM temperature distribution within a 2 $h^{-1}$cMpc wide region centered on the galaxy halo. The contours are drawn from the Mach number distribution along the line of sight and highlight the positions of various shock features. The position of the accretion shock is located towards the border of the shown maps (visible as transition of red to black), while the first and second internal shocks are located further in (mostly visible as transition from orange to red and yellow to orange in the left and middle panel). The middle panel shows all three shock features, the outer one and the two inner ones, very clearly. In the left panel the first internal shock and in the right panel the second internal shock are resembling the size of the observed ORCs, having a diameter of $\sim500$ kpc. In the middle and right panel, two rings inside each other could potentially shine in radio, while in the left panel the geometry could be interpreted as two overlapping elliptical shapes. The images (from left to right) show the halo at $z = 0.52$, $z = 0.43$ and $z = 0.27$, respectively.  
\label{fig:tmap}}
\end{figure}

\vspace{3cm}

\begin{figure*}[ht!]
\includegraphics[width=0.59\textwidth]{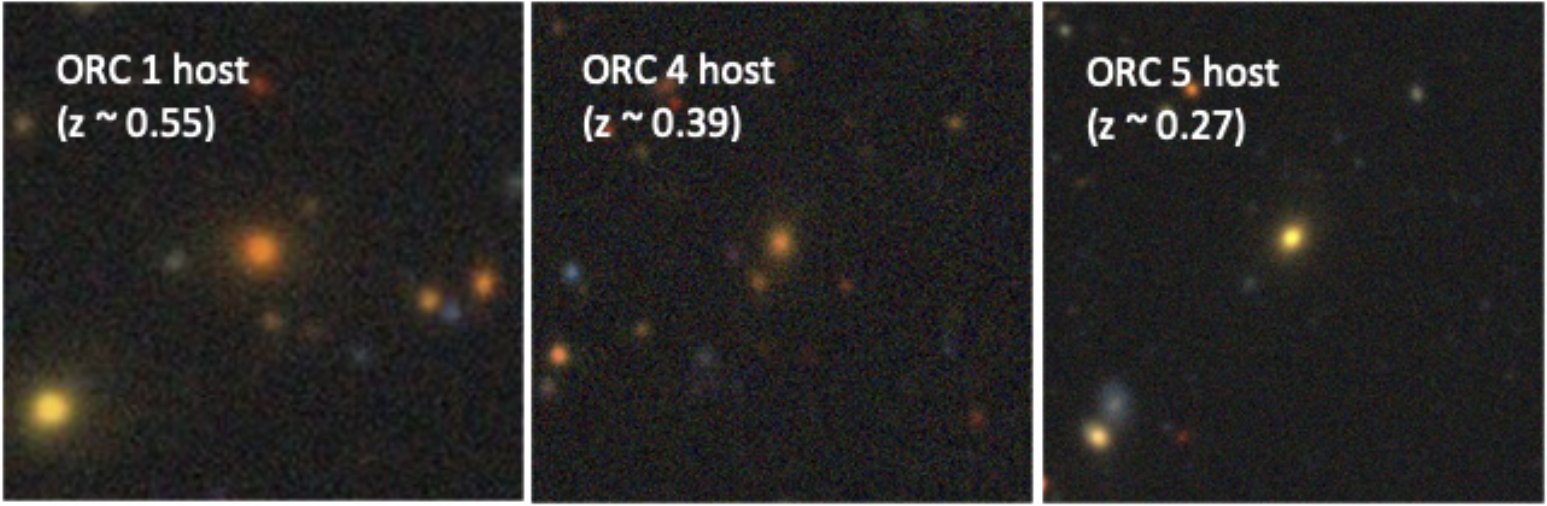}
\includegraphics[width=0.19\textwidth]{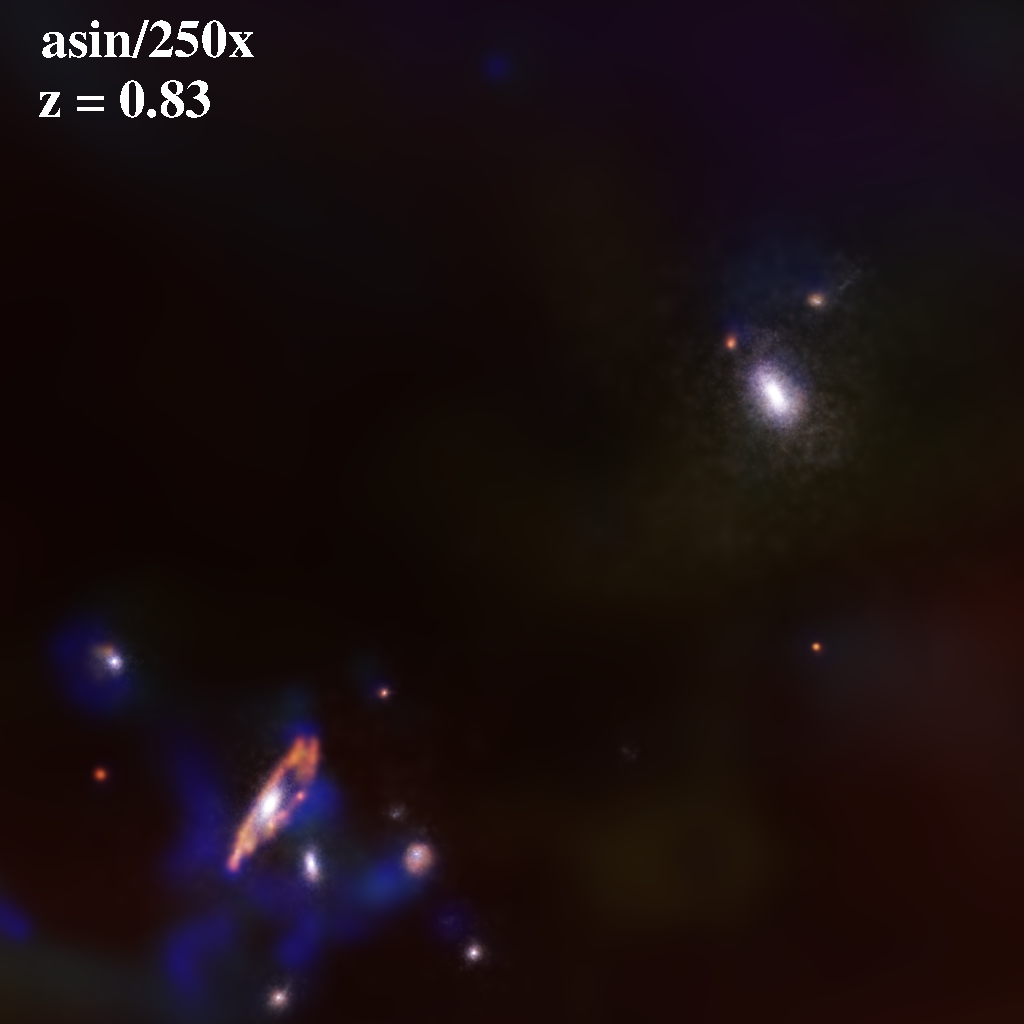}
\includegraphics[width=0.19\textwidth]{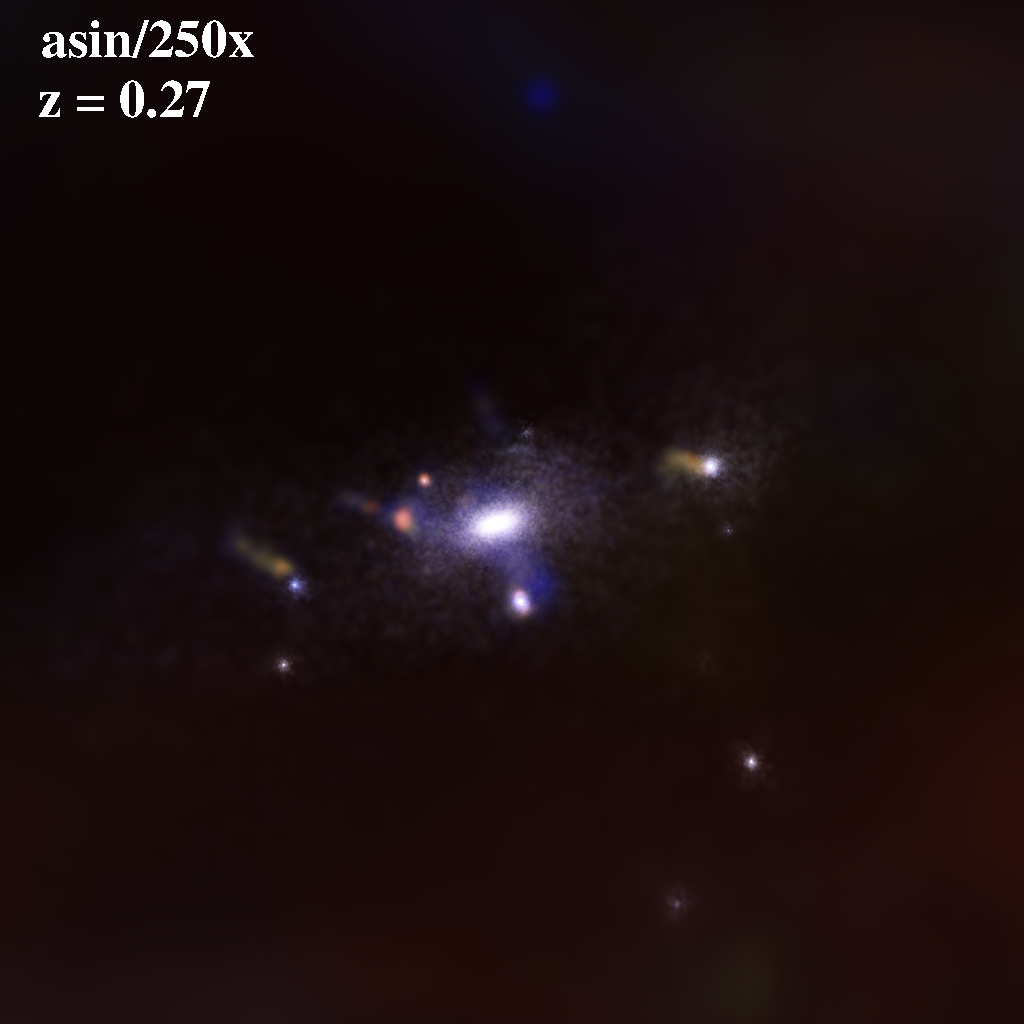}
\caption{The first three images (from the left) show DES optical color images centred on the likely host galaxies of the three single ORCs (see Table~2 in 
\citet{2021MNRAS.505L..11K}). --- The right two images made with \textsc{Splotch} \citep{2008NJPh...10l5006D} show the low-resolution simulation of the galaxy with galaxy formation physics on. The first shows the system at $z \approx0.83$, shortly before the first merger. The second shows the system at $z\approx0.27$, when the second shock has a size comparable with the observed diameter of the ORCs. Both are showing a region of 300~kpc a side.\label{fig:gal_map}}
\end{figure*}

\section{Internal shocks creating ORCs} \label{sec:internal_shocks}

In our simulated galaxy, a major merger event starting at $z\approx 0.7$ in the simulation is responsible for driving these two, subsequent shocks, which as shown before, both propagate far beyond the virial radius. During this propagation, they are causing roundish or horseshoe shaped features of similar size than the ORCs observed. Figure~\ref{fig:tmap} shows examples of temperature maps of the intergalactic medium (IGM), overlaid with the contours marking the positions of the detected shocks for a better demonstration of their shapes and sizes. Note that depending on how such shocks translate into radio emission, the geometry of the inner shock circle at $z\approx0.52$ shows significant asymmetry and also internal structure, which could appear as two folded circles, similar to those reported for recent MeerKAT observations of ORC~1 \citep{2022MNRAS.513.1300N}. In Section \ref{sec:3D_shocks} a three dimensional representation of the shock in the halo at $z=0.52$ is shown for better visualisation of the geometry. Therefore, internal shocks could also be responsible for such observed morphology, alternatively to a bipolar outflow as discussed in \citet{2022MNRAS.513.1300N}. Assuming that the acceleration of synchrotron bright electrons follows the same principles here as in galaxy clusters we can apply the standard model for radio relics \citep[see][and references therein]{2007MNRAS.375...77H}. There, the spectral index $\alpha$ is related to the radio Mach number $\mathcal{M}$ at injection through the simple relation $\alpha = (\mathcal{M}^2+1)/(\mathcal{M}^2-1) - \frac{1}{2}$. However, projection effects can have some influence on this relation, so for the integrated spectral index, the minus one halve typically vanishes \citep[see][for detailed discussion]{2021MNRAS.506..396W}. For ORC~1 this leads to a Mach number $\mathcal{M}\approx2.1-2.4$, in line with the Mach numbers inferred in \citet{2022MNRAS.513.1300N}, which also matches those of the internal, merger driven shocks seen in our simulations. This also indicates that if ORCs are produced by these internal shocks, a large variety of morphologies can be expected. Our galaxy candidate would have two periods, each spanning approx 500 million years, where the size of the internal shock structures matches the size of the observed ORCs. This would translate into at most 20 per cent of the time within the last 5--6 Gigayears. Similar to radio relics, projection effects and radial dependence of dissipated energy and magnetic field strength may play a significant role in the observability of ORCs \citep[see][for a discussion of these effects in galaxy clusters]{2012MNRAS.421.1868V}. This could lead to ORCs only being observable at a favorable distance to the central galaxy. This might contribute to the apparent phenomena that all discovered ORCs so far have similar observed sizes. Note also that sometimes internal shocks form inner and outer rings, as demonstrated in Figure~\ref{fig:tmap} (right panel). Depending on the actual, radial magnetic field profile within the CGM, these should be detectable in deeper observations.

\section{Formation of ORCs}
So far we have shown that the morphology of the internal shocks we found in our simulations matches the one observed for ORCs quite well. Specifically, in our case, we have two internal shocks  which pass through the halo on a very short timescale. This is due to the merging event at $z\approx0.7$ being a quite peculiar one. This merging event turns out to be a multiple merger, where the mass of the halo rises by a factor of three within $\approx 0.5$~Gyr. In the Appendix \ref{sec:mergertree} we show the detailed merger tree for a better understanding of the growth of this halo. Therefore, we expect this galaxy to be an early type galaxy. Indeed, the low-resolution simulation of this halo which includes star formation clearly shows that the galaxy undergoes a transformation from a disk like, star-forming system to a quiescent, early-type-like system after the merger. At the time when the morphology matches the ones of the observed ORCs (as at $z\approx0.45$), the galaxy appears as a standard, early-type galaxy.
Figure~\ref{fig:gal_map} shows the Dark Energy Survey \citep[DES;][]{2018ApJS..239...18A} optical colour images of the observed galaxies, as well as visualizations based on \textsc{Splotch} \citep{2008NJPh...10l5006D} of the low-resolution galaxy simulation including galaxy formation physics. One shows the system at $z\approx0.83$ before the merger starts, the other at $z\approx0.27$, when the second shock wave is an ORC candidate. Clear to see that before the merger, the simulation represents a late type, star-forming galaxy while after the merger the galaxy appears as a early type galaxy, with no sign of star-formation. The shown region spans in both cases 300~kpc in size.

From galaxy clusters with halo masses of $\approx10^{15}$~M$_\odot$ we know that such major merger events are the most energetic events in the Universe since the Big Bang, releasing energy of the order of $10^{64}$~erg and thereby powering shock-induced radio emission in form of radio relics with a typical power up to several times $10^{25}$~W/Hz at 1.4~GHz for the individual relics \citep{2012A&ARv..20...54F}. 
Therefore, a major merger event involving galaxies somewhat more massive than the Milky Way can power the observed ORCs, with an observed radio power of $\approx10^{23}$~W/Hz at 1.4~GHz.

\section{Expected frequency of ORCs}

\begin{figure}[t!]
\centering
\includegraphics[width=0.49\textwidth]{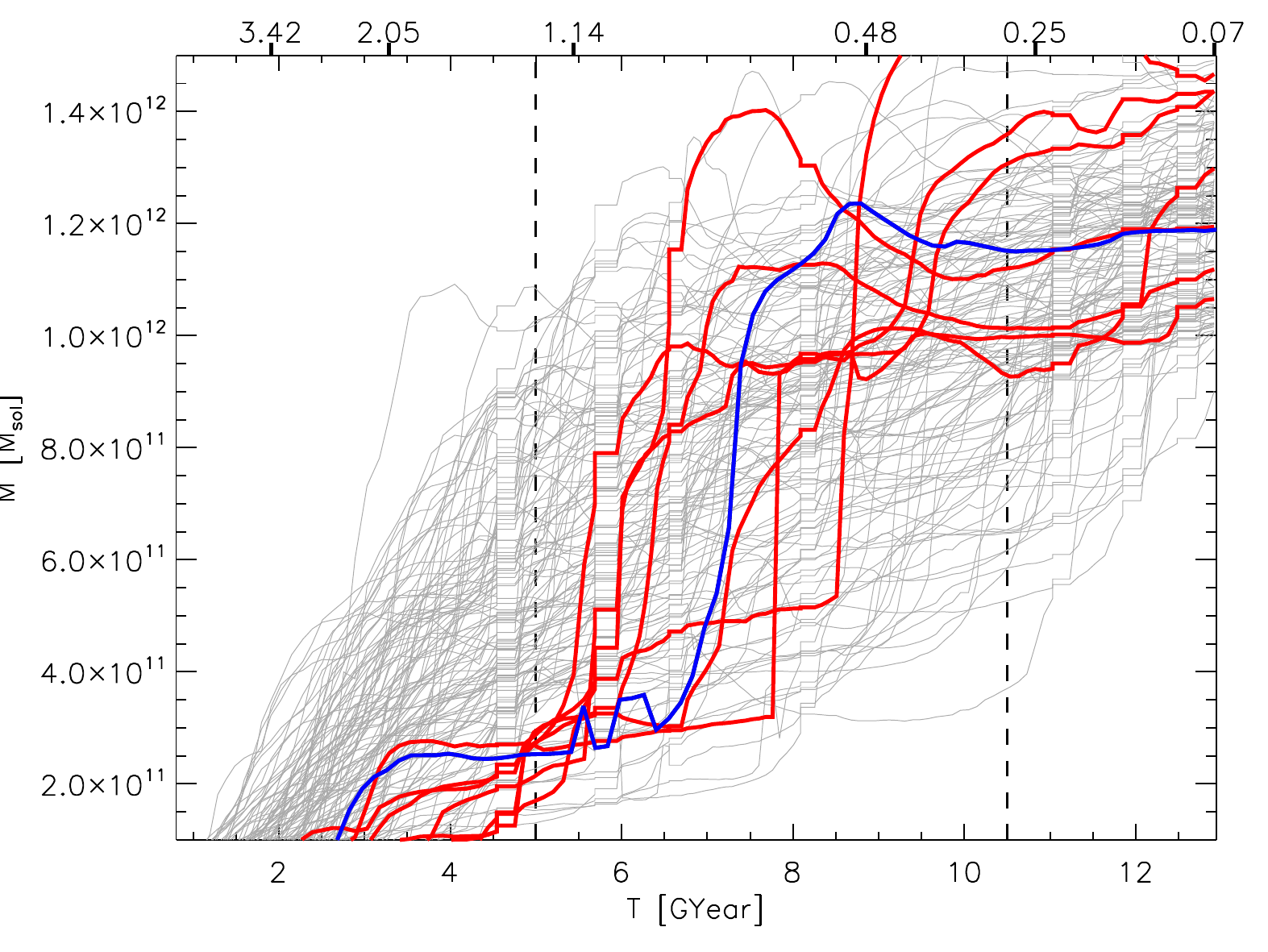}
\caption{We show the evolution of the mass of the main progenitor of the {\it asin/25000x} halo (thick blue line) compared to 123 galaxies from {\it Box4/uhr} of the {\it Magneticum} simulations (gray lines). Of those, 7 (red lines) are showing a similar shape of their formation history and a sudden, large increase of the virial mass, similar to our ORC candidate, within a reasonable time window given by the observations (indicated by the dashed lines).  \label{fig:main_all}}
\end{figure}

As the merger event which produces the morphology of the ORC candidate in our high-resolution galaxy simulation {\it asin/25000x} is quite unique, we can try to estimate the occurrence of such situations, which can lead to ORC candidates in larger scale, cosmological simulations. Here we used {\it Box4/uhr} from the {\it Magneticum} simulations to get a rough estimate of the expected occurrence of similar merger events, which involves similar, unusual multi-body merging processes. First we selected galaxies in the mass range $M_{\rm vir}$ between $1\times10^{12}$ and $1.5\times10^{12}$~M$_\odot$ to match the halo mass of our high-resolution simulation, resulting in a set of 123 galaxies. Then we obtained the frequency of similar merging events by finding mergers, where the mass of the main progenitor increases by a factor of three or more in a similarly short timescale. We identified seven such cases which could be classified to be similar to our candidate case. Figure~\ref{fig:main_all} shows the mass evolution of the main progenitor of the 123 selected galaxies, with the seven identified candidates shown as red lines. The blue line marks the mass evolution of the main progenitor of our high-resolution simulation. The dashed, vertical lines indicate the additional constraint of the time window, in which the major merger has to fall. In addition, this statistical set of heavy merging galaxies confirms the findings from the low-resolution simulation of the morphological type of our ORC candidates. Six out of the seven galaxies have zero star-formation rate and populate a region in the $j_*-M_*$ relation \citep[][]{2015ApJ...812...29T} which is related to early-type galaxies.

\begin{figure*}
\centering
\includegraphics[width=\textwidth]{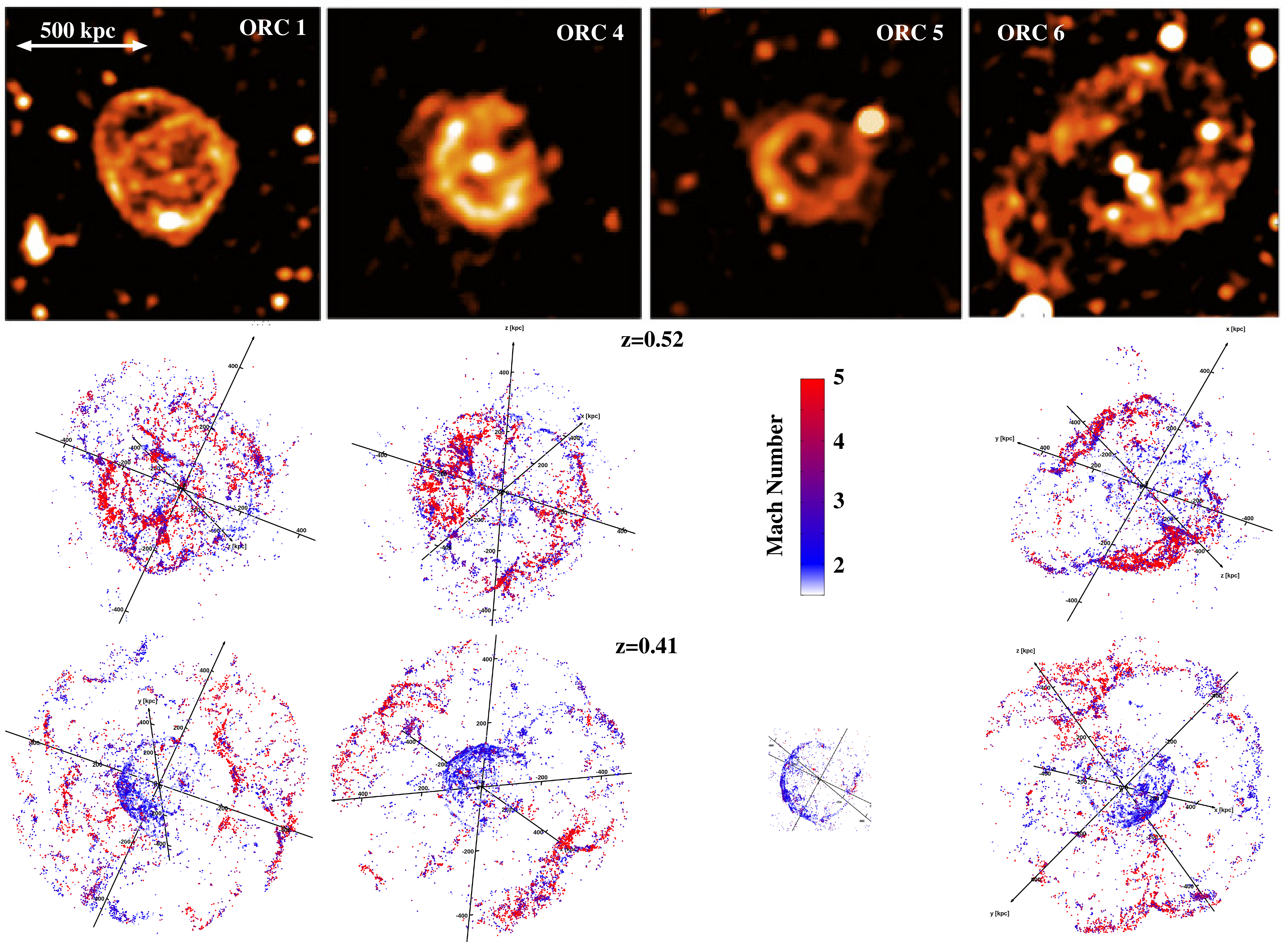}
\caption{The upper row shows the observed radio continuum emission of odd radio circles; from left to right: MeerKAT image of ORC~1 \citep{2022MNRAS.513.1300N}, GMRT image of ORC~4 \citep{2021PASA.38.3N}, and ASKAP images of ORC~5 \citep{2021MNRAS.505L..11K} and a newly discovered double ORC (Koribalski et al., in prep.). Each panel has a size of $\approx$ 1~Mpc $\times$ 1~Mpc. The lower rows shows 3D views of the shock structure under different viewing angles and at different times. We show the shocked SPH particles with $\mathcal{M}_s > 1.5$ color-coded according to their sonic Mach number. Time and viewing angle have been chosen to highlight morphological similarities between the structures of the internal shocks and the observed radio images. The 3D plots for interactively rotating the shock structures can be found on \url{http://www.magneticum.org/complements.html##COMPASS}}  
\label{fig:3D_shocks}
\end{figure*}

With this, we would expect that roughly 5 percent of galaxies in this mass range could at a certain point develop a radio ring feature. This upper limit has to be modulated down with the expected life time over which such feature is expected to be detectable and matches the observed geometry, which we can assume to be a window of 0.2 Gyr within the last 5 Gyr. The EMU Pilot Survey is spanning 270 square degree and ORCs are detected in a redshift range between $z\approx0.2$ and $z\approx0.5$ which corresponds to a volume of roughly 0.15~Gpc$^3$. From the typical mass functions predicted by cosmological simulations, we can estimate that the density of Milky Way-like galaxies and galaxies with mass of $\approx10^{13}$~M$_\odot$ is roughly $10^{-3}$ and $10^{-4}$ per Mpc$^3$, respectively. Therefore we expect that simulations would produce a few hundreds or some dozens of ORC candidates depending on the assumed halo masses. This is in line with the three single ORCs that have been confirmed so far. Note that in \citet{2022MNRAS.513.1300N} it was estimated that the ORC density should be around one per 0.05~Gpc$^3$, very similar to our estimate.

\section{3D Shock structures and appearance of ORCs} \label{sec:3D_shocks}

Our high-resolution simulation reveals internal shocks in the halo of galaxy mergers, as shown in Figure~\ref{fig:tmap}, that strongly resemble both the observed sizes and morphologies of ORCs. The 3D structures of these shock surfaces are visualised in Figure~\ref{fig:3D_shocks} for four different viewing angles at $z = 0.52$ and $z=0.43$. The viewing angle in the different columns are chosen to best reproduce the observed structures, like the left one, with many internal features, as observed in ORC~1 \citep{2022MNRAS.513.1300N}, or the right one, which highlights the gap between the two shock fronts, somewhat resembling a double ring or shell systems.
This demonstrates how the simulated system of shock surfaces viewed under different projections leads to various morphologies, from two-sided, to horseshoe or elliptical, with various sub-structures inside. In principal, this is quite similar to the situation of double relics in the outskirts of merging galaxy clusters. However, the shocks found here in the halos of galaxy mergers are more prominent at larger (relative to $R_{\mathrm vir}$) distances and appear more circular than their counterparts in clusters. 
Importantly, here we compare the appearance of just a single, forming halo in a cosmological context, and it is quite surprising that already this one case shows similarities to several, different observed systems. We expect that in a large cosmological sample of galaxies there will be an even broader spectrum of geometries and therefore morphologies produced in such merging events. 

\begin{figure}[th!]
\includegraphics[width=0.49\textwidth]{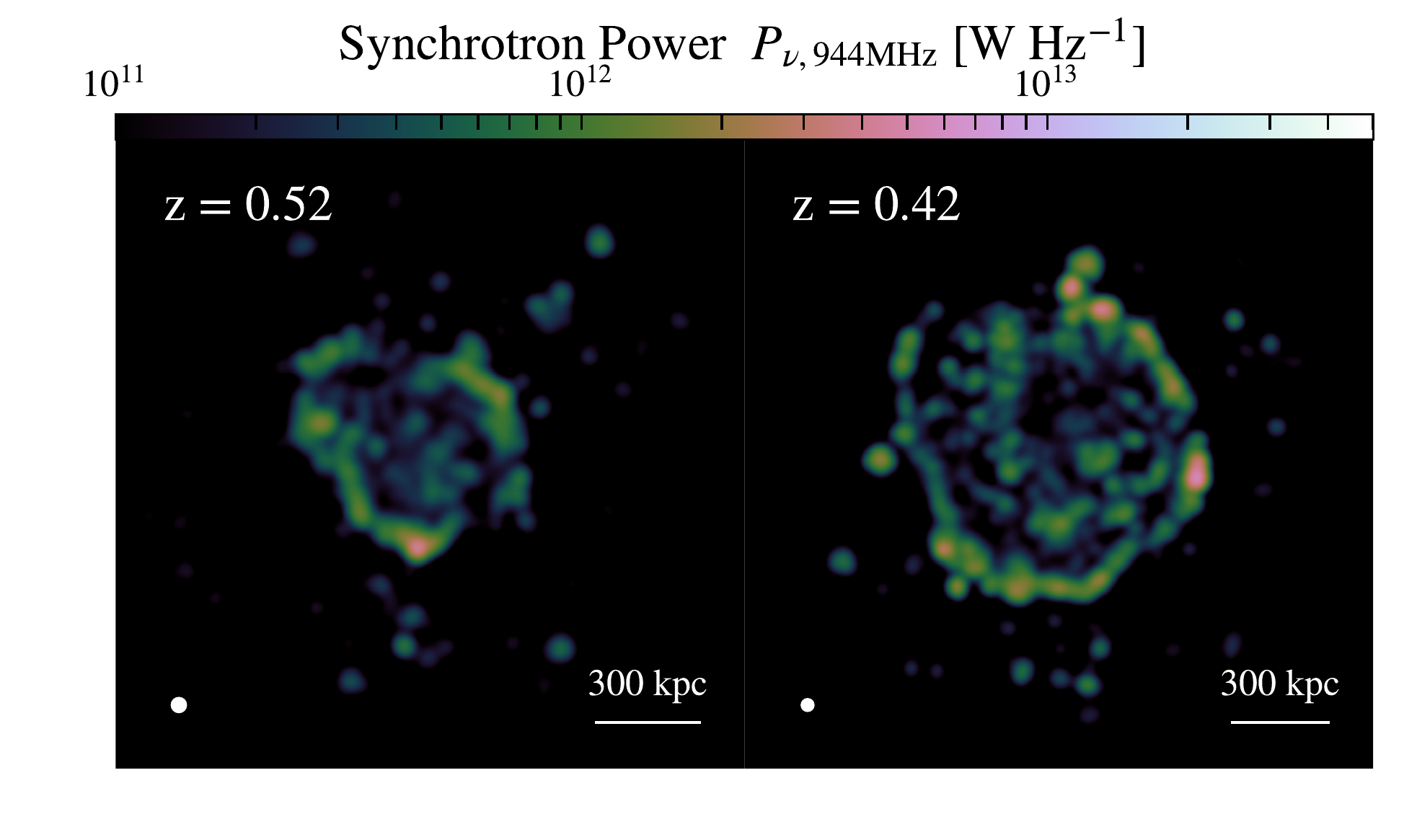}
\caption{Two example for the radio synchrotron emission at 944~MHz of the simulated system at $z=0.52$ and $z=0.42$. The images are smoothed with a 13 arcsec beam, corresponding to the observations. The beam size is indicated as a small white circle in the lower left corner.
\label{fig:radio_map}}
\end{figure}

\section{Estimation of the radio emission}

As already pointed out in \citet{2022MNRAS.513.1300N}, shocks with Mach numbers $\mathcal{M} \approx 2-3$ could cause the observed synchrotron emission. For this assumption to hold, the energy density of the CR electrons in the synchrotron emitting  shell has to be of order $10^4$~eV~m$^{-3}$, corresponding to roughly 1\% of the energy density of the CMB at this redshift. However, the thermal energy density at these distances from the galaxy, for typical temperatures and densities, is of the same order and therefore might challenge our  current understanding of diffuse shock acceleration (DSA) from a thermal pool.  

Here, we estimate the radio luminosity of our simulated ORC candidates by modeling the cosmic ray (CR) electron population through our Fokker-Planck solver \textsc{Crescendo} \citep[][]{2022arXiv220705087B}. For this purpose we stored $\approx1000$ snapshots of the {\it asin/25000x} simulation to trace the shock structures in detail. We used the efficiencies from \citet{2019ApJ...883...60R} to model the acceleration of CR protons at shocks via diffuse shock acceleration (DSA) and employ an electron-to-proton injection ratio of $K_{\rm ep} = 0.01$.
For the treatment of the cosmic ray electrons we focus on radiative cooling by inverse Compton scattering and ignore adiabatic changes, as they are found to be of reduced relevance in the case of studying shock fronts \citep[see][for a discussion]{2021arXiv211200023W}. The cosmic ray electron spectrum is represented in the range $p \in [10^2, 10^5] (m_e c)^{-1}$ with 8 bins/dex.
As in the on-the-fly implementation, we subcycle the solver for particles where the time between outout snapshots $\Delta t_{\mathrm{snapshot}} > 0.1 \: \tau_{\mathrm{IC}}(p_{\mathrm{cut}})$ where $\tau_{\mathrm{IC}}(p_{\mathrm{cut}})$ is the cooling time of the electrons at the spectral cutoff due to inverse Compton scattering.

In Figure~\ref{fig:radio_map} we shows two examples of the predicted synchrotron emission from our system at $z=0.52$ and $z=0.42$. Also here, even the projected radio emission follows quite closely the ones observed from the shocks itself. In this case, there are noticeable similarities to the different cases of observed ORC systems. The size of the emitting structures reaches a diameter of $\approx$ 500 kpc (left pane) and even larger sizes (right panel), with some remaining emission structures further in, which could be interpreted at internal structures of a double ring system as in ORC~1 or ORC~6. Increasing the dynamic range in Figure~\ref{fig:radio_map} to show darker structures reveals a second ORC around the inner one. This is caused by a previous merger event and its still outwards traveling shock, as already seen in the temperature profiles. The observability of this second layer of ORCs would however strongly depend on the CR acceleration physics and the magnetic field this far out in the CGM.

\begin{figure}[th!]
\centering
\includegraphics[width=0.48\textwidth]{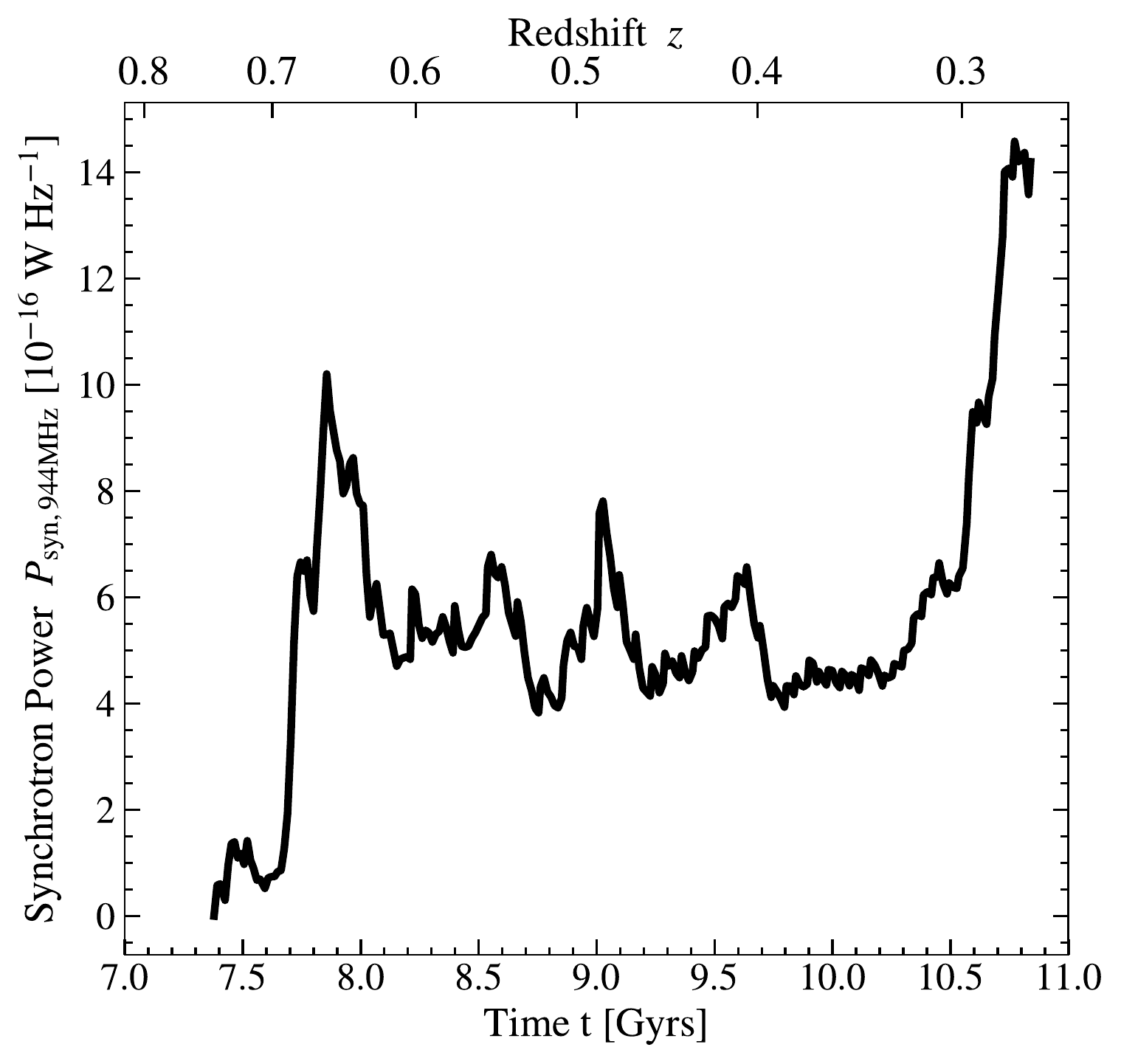}
\caption{Total synchrotron brightness of the simulated ORC within a radius of $r < 500 h^{-1} c$Mpc for our chosen CR acceleration efficiencies models. We use a constant magnetic field strength of $B = 5 \mu \mathrm{G}$ for all synchrotron emitting particles. Note that the observed synchrotron power at 944~MHz for ORC~5 is significant larger, e.g. $\approx 5.5\times10^{23}$ W\,Hz$^{-1}$ \citep{2021MNRAS.505L..11K}. 
\label{fig:total_radio_emission}}
\end{figure}

Figure~\ref{fig:total_radio_emission} shows the evolution of the total synchrotron brightness of these structures for the choice of our DSA efficiency model. Very clear to see that such emission would appear very variable, changing by a factor of a view over timescales on Gyr and below. The emission only starts to get significant, after the first merger shocks are initiated. However, as discussed before, our simplistic models fail to reproduce the observed total luminosity, due to the low, thermal energy content of the CGM at these distances. Therefore it is difficult to state exactly for which time periods the shock structures would light up and could produce the observed radio emission of ORCs.

Our direct application of DSA strongly depends on many assumptions which are beyond both, the currently underlying hydro-dynamical simulation as well as our cosmic ray modeling. Among them, the obtained Mach number is crucial for the ability to accelerate cosmic ray electrons, and therefore the obtained radio brightness and its smoothness is very uncertain and should be only taken as an indication of the expected morphology or the principle variation of the time evolution. Despite our generally very high resolution for such kind of simulations, the detection of the shocks in the outer parts is still quite limited by numerical effects, as already noticed in the shock maps. We expect that processes like cooling and feedback will have a noticeable impact on the CGM at these distances and therefore the involved Mach numbers will change. Going from a halo with a virial mass of $10^{12}$~M$_\odot$ to a halo with mass of $10^{13}$~M$_\odot$ can potentially increase the radio luminosity, as observed through the scaling relations for radio relics in galaxy clusters.  Furthermore, we are neglecting effects like turbulent re-acceleration and maybe most importantly, fossil cosmic ray electron populations from galactic outflows and AGN activity, which can be more efficiently re-accelerated. Therefore we stress that the presented radio emission should only be interpreted as an indication that the underlying shock structures from the formation process of galactic halos may explain the formation of ORCs.

\section{Detectability of the shocks in X-rays}

\begin{figure}[th!]
\centering
\includegraphics[width=0.48\textwidth]{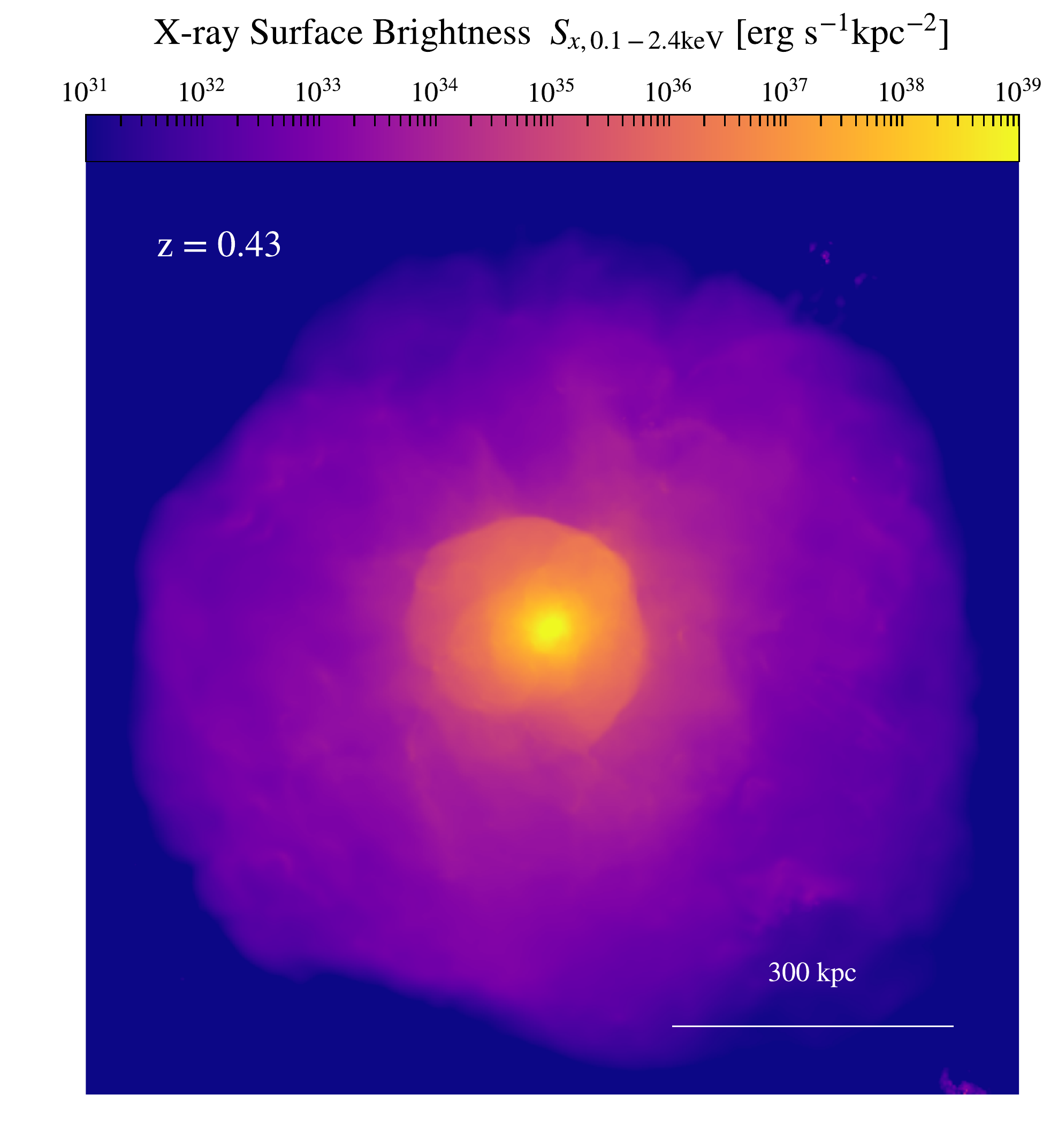}
\caption{Shown is the intrinsic surface brightness of the simulated galaxy at $z=0.43$ using an oversimplified, thermal model for the expected X-ray emission of the CGM assuming primordial composition. The location of the shocks is clearly visible as significant jumps in the surface brightness. See text for discussion on the expected level of surface brightness and the detectability with future X-ray instruments. 
\label{fig:xray_sb}}
\end{figure}

In galaxy clusters, such shocks are nowadays routinely detected in X-ray observations as surface brightness or temperature jumps \citep[see][for examples]{2007PhR...443....1M}. As shown in Figure~\ref{fig:xray_sb}, also in our ORC candidate these shocks are featuring steep surface brightness features. However, detecting such features within the CGM, especially at these distances from the stellar body of the galaxies is out of range for current instruments. Furthermore, the actual brightness of the CGM at these densities and temperatures is very uncertain, as it depends not only on highly unknown effects of galaxy formation physics \citep[see][]{2022arXiv221109827K}, but also on the metallicity of the CGM, and will be strongly moderated by effects like resonant scattering of the X-ray photons. In sum, all these effects can alter the expected surface brightness by more than an order of magnitude and therefore the absolute values in Figure~\ref{fig:xray_sb} should not be taken as face values. Nevertheless, future X-ray missions like the Line Emission Mapper \citep[LEM,][]{2022arXiv221109827K}.
should be able to detect such emission of the CGM, where the mean over-densities compared to the cosmic value is $\approx 100$ and the temperatures are slightly below 0.1 keV, as can be inferred from current sensitivity predictions (Churazov et al., in prep.). This is especially interesting, as a recent discovery in ASKAP surveys indicates such features to be present also in local galaxies at $z \approx 0.05$ (The Cloverleaf, Koribalski et al., in prep.), which would allow to distinguish the line emission from elements like Oxygen (O\,{\sc vii} and O\,{\sc viii}) from the galactic foreground. 

\section{Conclusions}

We present the results of non-radiative, very high-resolution simulations of a Milky Way mass galactic halo where we applied an on-the-fly shock finder.
We show that similar to what was found for merger-accelerated accretion shocks (MA-shocks) in galaxy clusters \citep[see][]{2020MNRAS.494.4539Z}, the internal shocks from the formation of the halo are colliding with the accretion shock and pushing it back to a distance of several times the virial radius. In some cases, subsequent internal shocks driven by merger events are able to produce very complex shock structures, prominently propagating even up to the virial radius and beyond. These structures could be the origin of a newly detected radio emission feature of galaxies called {\it odd radio circles} (ORCs), see \citet{2021PASA.38.3N}. The best matching morphology found in our galaxy simulation originates from a very-extreme merger event, where the final halo mass of the galaxy increases by a factor of three due to multiple merger components. By identifying similar extreme merger cases in a cosmological simulation, we deduced that only 5\% of galaxies undergo such dramatic events, which would be generally in line with the rare number of observations of such ORCs. Here, the lower abundance of more massive halos with virial masses around $10^{13}$~M$_\odot$ would even better match the currently observed frequency of ORCs, which lines up with halo masses expected for the observed stellar masses of several $10^{11}$~M$_\odot$. This scenario also predicts that the galaxies showing these radio emission phenomena should be predominately quiescent galaxies, which is confirmed by a lower-resolution simulation of our target galaxy including cooling and star formation and lines up with the observational findings for the small number of ORCs detected so far. Furthermore, our scenario of internal shocks also agrees well with the radio Mach number of $\mathcal{M} \approx 2.1-2.4$ inferred from observations of ORC~1.

Following a spectral distribution of cosmic ray electrons which are produced via diffuse shock acceleration at these shocks with \textsc{Crescendo} \citep[][]{2022arXiv220705087B} we verified the morphology and the transient nature of the expected radio emission. However, given the over-simplified treatment of the complex processes involved, applying  direct shock acceleration for the CR electrons at these shocks was not yet able to reproduce the magnitude of the observed radio emission.

Such shocks should be detectable as low surface brightness features with future X-ray missions like LEM.

In summary, we find that internal merger shocks are an interesting new candidate to explain the newly discovered odd circular radio emission around some galaxies by matching several of the observed properties. More detailed modeling of the cosmic ray electron component causing this radio emission will be needed to better understand the formation of ORCs.

\begin{acknowledgments}
We want to thank Michelle Lochner for providing an image of the recently discovered object named SAURON, produced using the data presented in \citep{2022arXiv221102062L}. KD, LMB and MV acknowledge support by the Deutsche Forschungsgemeinschaft (DFG, German Research Foundation) under Germanys Excellence Strategy -- EXC-2094 -- 390783311. KD acknowledges support for the COMPLEX project from the European Research Council (ERC) under the European Union’s Horizon 2020 research and innovation program grant agreement ERC-2019-AdG 882679. The calculations were carried out at the Leibniz Supercomputer Center (LRZ) under the projects pr83li and pr86re. UPS is supported by the Simons Foundation through a Flatiron Research Fellowship at the Center for Computational Astrophysics of the Flatiron Institute. The Flatiron Institute is supported by the Simons Foundation. MV is supported by the Alexander von Humboldt Stiftung and the Carl Friedrich von Siemens Stiftung. We make use of public archival images from the Dark Energy Survey (DES) and acknowledge the institutions listed on \url{https://www.darkenergysurvey.org/the-des-project/data-access/}. \\ 
\end{acknowledgments}

\software{\textsc{P-Gadget3} \citep{2005MNRAS.364.1105S,2016MNRAS.455.2110B},  
          \textsc{Smac} \citep{2005MNRAS.363...29D},
          \textsc{SubFind} \citep{2001MNRAS.328..726S,2009MNRAS.399..497D},
          \textsc{Splotch} \citep{2008NJPh...10l5006D},
          Julia \citep{bezanson2017julia}, GadgetIO.jl \citep[][]{GadgetIO},
          Matplotlib \citep{Hunter2007}
          }

\appendix

\section{Appendix information}

\subsection{Detailed formation history} \label{sec:mergertree}

\begin{figure}
\includegraphics[width=1.0\textwidth]{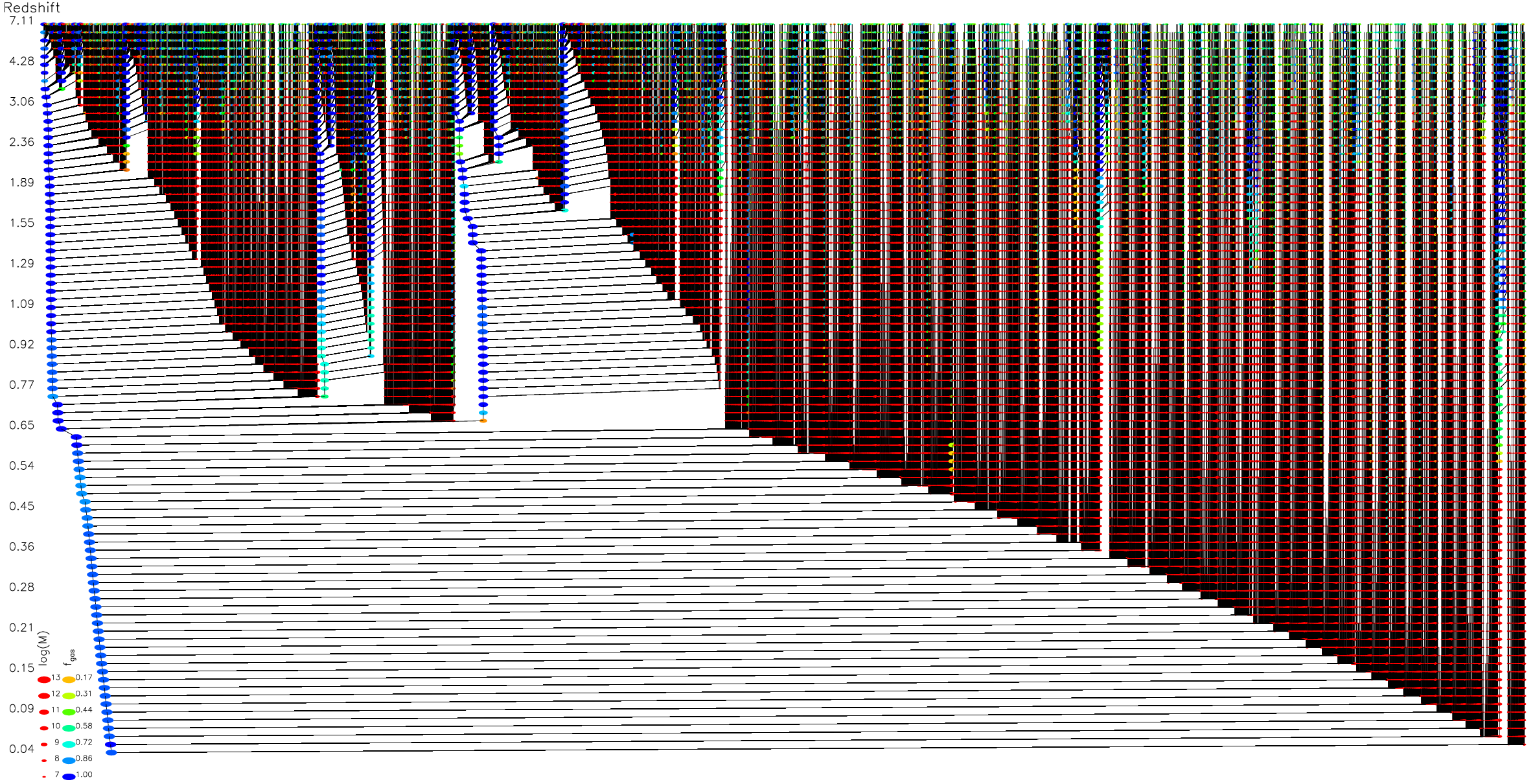}
\caption{Merger tree of our high-resolution galaxy simulations, showing the triple merger event at $z = 0.7$ and $z=0.65$ leading to internal shocks which resemble the ORC geometry at $z \sim 0.45$. See text for details. \label{fig:merger_tree}}
\end{figure}

In Figure~\ref{fig:merger_tree} we show the merger tree based on \textsc{SubFind} \citep{2001MNRAS.328..726S,2009MNRAS.399..497D}, constructed for our high-resolution galaxy simulation. On the vertical axis, the redshift is given, staring from $z=7.11$ in the top and goes down to $z=0$ at the bottom, showing only 90 out of the originally 1000 output times of the simulation. Each of the colored circle represent a galaxy at each time, where the size of the circles are logarithmic scaled with the current halo mass and color is indicating the gas mass fraction in units of the cosmic mean, as indicated in the legend. We restricted the visualisation of the tree to halos which are resolved with at least 100 particles, leading to $\approx$33000 progenitors shown in the figure on the horizontal axis. The black lines connect where each galaxy is ending up at the next time, so vertical lines indicate the evolution of individual galaxies and horizontal lines indicate when galaxies are finally merged into other galaxies. The left branch shows the evolution of the so called main progenitor, e.g. the most massive progenitor at each time which ends up being part of the final galaxy. Clear to see is that the merger events which take place between $z\approx0.7$ and $z\approx0.65$\footnote{These are the times when the halos are touching. The galaxies (which are the central parts) will merge significantly later and therefore also the shocks which are driven by the core passage will be initiated significantly later.} are quite unusual. Here, almost simultaneously three approximately equal halos are merging, as can be seen from their individually rich and complex history, manifesting in their relatively large number of own leaves in the tree. This happens in a very short timescale, which then also drives the two independent shock waves, which could resemble the ORC morphology as discussed in the main text. The mass of the halo also basically increased by a factor of three by this event. As a side note, it is also nicely visible how all the individual galaxies are losing their diffuse gas in interaction with the host system before finally merging, as can be seen from the appearance of the vertical color gradients before the horizontal lines indicating the merging into the final galaxy.  

\bibliography{paper_sorgs}{}

\begin{thebibliography}{}
\expandafter\ifx\csname natexlab\endcsname\relax\def\natexlab#1{#1}\fi
\providecommand{\url}[1]{\href{#1}{#1}}
\providecommand{\dodoi}[1]{doi:~\href{http://doi.org/#1}{\nolinkurl{#1}}}
\providecommand{\doeprint}[1]{\href{http://ascl.net/#1}{\nolinkurl{http://ascl.net/#1}}}
\providecommand{\doarXiv}[1]{\href{https://arxiv.org/abs/#1}{\nolinkurl{https://arxiv.org/abs/#1}}}

\bibitem[{{Abbott} {et~al.}(2018){Abbott}, {Abdalla}, {Allam}, {Amara},
  {Annis}, {Asorey}, {Avila}, {Ballester}, {Banerji}, {Barkhouse}, {Baruah},
  {Baumer}, {Bechtol}, {Becker}, {Benoit-L{\'e}vy}, {Bernstein}, {Bertin},
  {Blazek}, {Bocquet}, {Brooks}, {Brout}, {Buckley-Geer}, {Burke}, {Busti},
  {Campisano}, {Cardiel-Sas}, {Carnero Rosell}, {Carrasco Kind}, {Carretero},
  {Castander}, {Cawthon}, {Chang}, {Chen}, {Conselice}, {Costa}, {Crocce},
  {Cunha}, {D'Andrea}, {da Costa}, {Das}, {Daues}, {Davis}, {Davis}, {De
  Vicente}, {DePoy}, {DeRose}, {Desai}, {Diehl}, {Dietrich}, {Dodelson},
  {Doel}, {Drlica-Wagner}, {Eifler}, {Elliott}, {Evrard}, {Farahi}, {Fausti
  Neto}, {Fernandez}, {Finley}, {Flaugher}, {Foley}, {Fosalba}, {Friedel},
  {Frieman}, {Garc{\'\i}a-Bellido}, {Gaztanaga}, {Gerdes}, {Giannantonio},
  {Gill}, {Glazebrook}, {Goldstein}, {Gower}, {Gruen}, {Gruendl}, {Gschwend},
  {Gupta}, {Gutierrez}, {Hamilton}, {Hartley}, {Hinton}, {Hislop}, {Hollowood},
  {Honscheid}, {Hoyle}, {Huterer}, {Jain}, {James}, {Jeltema}, {Johnson},
  {Johnson}, {Kacprzak}, {Kent}, {Khullar}, {Klein}, {Kovacs}, {Koziol},
  {Krause}, {Kremin}, {Kron}, {Kuehn}, {Kuhlmann}, {Kuropatkin}, {Lahav},
  {Lasker}, {Li}, {Li}, {Liddle}, {Lima}, {Lin}, {L{\'o}pez-Reyes}, {MacCrann},
  {Maia}, {Maloney}, {Manera}, {March}, {Marriner}, {Marshall}, {Martini},
  {McClintock}, {McKay}, {McMahon}, {Melchior}, {Menanteau}, {Miller},
  {Miquel}, {Mohr}, {Morganson}, {Mould}, {Neilsen}, {Nichol}, {Nogueira},
  {Nord}, {Nugent}, {Nunes}, {Ogando}, {Old}, {Pace}, {Palmese},
  {Paz-Chinch{\'o}n}, {Peiris}, {Percival}, {Petravick}, {Plazas}, {Poh},
  {Pond}, {Porredon}, {Pujol}, {Refregier}, {Reil}, {Ricker}, {Rollins},
  {Romer}, {Roodman}, {Rooney}, {Ross}, {Rykoff}, {Sako}, {Sanchez}, {Sanchez},
  {Santiago}, {Saro}, {Scarpine}, {Scolnic}, {Serrano}, {Sevilla-Noarbe},
  {Sheldon}, {Shipp}, {Silveira}, {Smith}, {Smith}, {Smith}, {Soares-Santos},
  {Sobreira}, {Song}, {Stebbins}, {Suchyta}, {Sullivan}, {Swanson}, {Tarle},
  {Thaler}, {Thomas}, {Thomas}, {Troxel}, {Tucker}, {Vikram}, {Vivas},
  {Walker}, {Wechsler}, {Weller}, {Wester}, {Wolf}, {Wu}, {Yanny}, {Zenteno},
  {Zhang}, {Zuntz}, {DES Collaboration}, {Juneau}, {Fitzpatrick}, {Nikutta},
  {Nidever}, {Olsen}, {Scott}, \& {NOAO Data Lab}}]{2018ApJS..239...18A}
{Abbott}, T.~M.~C., {Abdalla}, F.~B., {Allam}, S., {et~al.} 2018, \apjs, 239,
  18, \dodoi{10.3847/1538-4365/aae9f0}

\bibitem[{{Beck} {et~al.}(2016{\natexlab{a}}){Beck}, {Dolag}, \&
  {Donnert}}]{2016MNRAS.458.2080B}
{Beck}, A.~M., {Dolag}, K., \& {Donnert}, J.~M.~F. 2016{\natexlab{a}}, \mnras,
  458, 2080, \dodoi{10.1093/mnras/stw487}

\bibitem[{{Beck} {et~al.}(2016{\natexlab{b}}){Beck}, {Murante}, {Arth},
  {Remus}, {Teklu}, {Donnert}, {Planelles}, {Beck}, {F{\"o}rster}, {Imgrund},
  {Dolag}, \& {Borgani}}]{2016MNRAS.455.2110B}
{Beck}, A.~M., {Murante}, G., {Arth}, A., {et~al.} 2016{\natexlab{b}}, \mnras,
  455, 2110, \dodoi{10.1093/mnras/stv2443}

\bibitem[{{Bennett} \& {Sijacki}(2020)}]{2020MNRAS.499..597B}
{Bennett}, J.~S., \& {Sijacki}, D. 2020, \mnras, 499, 597,
  \dodoi{10.1093/mnras/staa2835}

\bibitem[{Bezanson {et~al.}(2017)Bezanson, Edelman, Karpinski, \&
  Shah}]{bezanson2017julia}
Bezanson, J., Edelman, A., Karpinski, S., \& Shah, V.~B. 2017, SIAM review, 59,
  65.
\newblock \url{https://doi.org/10.1137/141000671}

\bibitem[{{Bolton} {et~al.}(2012){Bolton}, {Schlegel}, {Aubourg}, {Bailey},
  {Bhardwaj}, {Brownstein}, {Burles}, {Chen}, {Dawson}, {Eisenstein}, {Gunn},
  {Knapp}, {Loomis}, {Lupton}, {Maraston}, {Muna}, {Myers}, {Olmstead},
  {Padmanabhan}, {P{\^a}ris}, {Percival}, {Petitjean}, {Rockosi}, {Ross},
  {Schneider}, {Shu}, {Strauss}, {Thomas}, {Tremonti}, {Wake}, {Weaver}, \&
  {Wood-Vasey}}]{2012AJ....144..144B}
{Bolton}, A.~S., {Schlegel}, D.~J., {Aubourg}, {\'E}., {et~al.} 2012, \aj, 144,
  144, \dodoi{10.1088/0004-6256/144/5/144}

\bibitem[{{Bonafede} {et~al.}(2011){Bonafede}, {Dolag}, {Stasyszyn}, {Murante},
  \& {Borgani}}]{2011MNRAS.418.2234B}
{Bonafede}, A., {Dolag}, K., {Stasyszyn}, F., {Murante}, G., \& {Borgani}, S.
  2011, \mnras, 418, 2234, \dodoi{10.1111/j.1365-2966.2011.19523.x}

\bibitem[{{Bortolas}(2022)}]{2022MNRAS.511.2885B}
{Bortolas}, E. 2022, \mnras, 511, 2885, \dodoi{10.1093/mnras/stac262}

\bibitem[{{B{\"o}ss} {et~al.}(2023){B{\"o}ss}, {Steinwandel}, {Dolag}, \&
  {Lesch}}]{2022arXiv220705087B}
{B{\"o}ss}, L.~M., {Steinwandel}, U.~P., {Dolag}, K., \& {Lesch}, H. 2023,
  \mnras, 519, 548, \dodoi{10.1093/mnras/stac3584}

\bibitem[{Böss \& Valenzuela(2022)}]{GadgetIO}
Böss, L.~M., \& Valenzuela, L.~M. 2022, LudwigBoess/GadgetIO.jl: v0.6.0,
  v0.6.0,  Zenodo, \dodoi{10.5281/zenodo.7053305}

\bibitem[{{Cox} {et~al.}(2006){Cox}, {Jonsson}, {Primack}, \&
  {Somerville}}]{2006MNRAS.373.1013C}
{Cox}, T.~J., {Jonsson}, P., {Primack}, J.~R., \& {Somerville}, R.~S. 2006,
  \mnras, 373, 1013, \dodoi{10.1111/j.1365-2966.2006.11107.x}

\bibitem[{{Cox} {et~al.}(2004){Cox}, {Primack}, {Jonsson}, \&
  {Somerville}}]{2004ApJ...607L..87C}
{Cox}, T.~J., {Primack}, J., {Jonsson}, P., \& {Somerville}, R.~S. 2004, \apjl,
  607, L87, \dodoi{10.1086/421905}

\bibitem[{{Dolag} {et~al.}(2009){Dolag}, {Borgani}, {Murante}, \&
  {Springel}}]{2009MNRAS.399..497D}
{Dolag}, K., {Borgani}, S., {Murante}, G., \& {Springel}, V. 2009, \mnras, 399,
  497, \dodoi{10.1111/j.1365-2966.2009.15034.x}

\bibitem[{{Dolag} {et~al.}(2005){Dolag}, {Hansen}, {Roncarelli}, \&
  {Moscardini}}]{2005MNRAS.363...29D}
{Dolag}, K., {Hansen}, F.~K., {Roncarelli}, M., \& {Moscardini}, L. 2005,
  \mnras, 363, 29, \dodoi{10.1111/j.1365-2966.2005.09452.x}

\bibitem[{{Dolag} {et~al.}(2008){Dolag}, {Reinecke}, {Gheller}, \&
  {Imboden}}]{2008NJPh...10l5006D}
{Dolag}, K., {Reinecke}, M., {Gheller}, C., \& {Imboden}, S. 2008, New Journal
  of Physics, 10, 125006, \dodoi{10.1088/1367-2630/10/12/125006}

\bibitem[{{Feretti} {et~al.}(2012){Feretti}, {Giovannini}, {Govoni}, \&
  {Murgia}}]{2012A&ARv..20...54F}
{Feretti}, L., {Giovannini}, G., {Govoni}, F., \& {Murgia}, M. 2012, \aapr, 20,
  54, \dodoi{10.1007/s00159-012-0054-z}

\bibitem[{{Gupta} {et~al.}(2022){Gupta}, {Huynh}, {Norris}, {Wang}, {Hopkins},
  {Andernach}, {Koribalski}, \& {Galvin}}]{2022arXiv220813997G}
{Gupta}, N., {Huynh}, M., {Norris}, R.~P., {et~al.} 2022, arXiv e-prints,
  arXiv:2208.13997.
\newblock \doarXiv{2208.13997}

\bibitem[{{Hani} {et~al.}(2018){Hani}, {Sparre}, {Ellison}, {Torrey}, \&
  {Vogelsberger}}]{2018MNRAS.475.1160H}
{Hani}, M.~H., {Sparre}, M., {Ellison}, S.~L., {Torrey}, P., \& {Vogelsberger},
  M. 2018, \mnras, 475, 1160, \dodoi{10.1093/mnras/stx3252}

\bibitem[{{Hoeft} \& {Br{\"u}ggen}(2007)}]{2007MNRAS.375...77H}
{Hoeft}, M., \& {Br{\"u}ggen}, M. 2007, \mnras, 375, 77,
  \dodoi{10.1111/j.1365-2966.2006.11111.x}

\bibitem[{Hunter(2007)}]{Hunter2007}
Hunter, J.~D. 2007, Computing in Science \& Engineering, 9, 90,
  \dodoi{10.1109/MCSE.2007.55}

\bibitem[{{Kirillov} \& {Savelova}(2020)}]{2020EPJC...80..810K}
{Kirillov}, A.~A., \& {Savelova}, E.~P. 2020, European Physical Journal C, 80,
  810, \dodoi{10.1140/epjc/s10052-020-8395-7}

\bibitem[{{Koribalski} {et~al.}(2021){Koribalski}, {Norris}, {Andernach},
  {Rudnick}, {Shabala}, {Filipovi{\'c}}, \& {Lenc}}]{2021MNRAS.505L..11K}
{Koribalski}, B.~S., {Norris}, R.~P., {Andernach}, H., {et~al.} 2021, \mnras,
  505, L11, \dodoi{10.1093/mnrasl/slab041}

\bibitem[{{Kraft} {et~al.}(2022){Kraft}, {Markevitch}, {Kilbourne}, \& {the LEM
  Team}}]{2022arXiv221109827K}
{Kraft}, R., {Markevitch}, M., {Kilbourne}, C., \& {the LEM Team}. 2022, arXiv
  e-prints, arXiv:2211.09827.
\newblock \doarXiv{2211.09827}

\bibitem[{{Lochner} {et~al.}(2022){Lochner}, {Rudnick}, {Heywood}, {Knowles},
  \& {Shabala}}]{2022arXiv221102062L}
{Lochner}, M., {Rudnick}, L., {Heywood}, I., {Knowles}, K., \& {Shabala}, S.~S.
  2022, arXiv e-prints, arXiv:2211.02062.
\newblock \doarXiv{2211.02062}

\bibitem[{{Markevitch} \& {Vikhlinin}(2007)}]{2007PhR...443....1M}
{Markevitch}, M., \& {Vikhlinin}, A. 2007, \physrep, 443, 1,
  \dodoi{10.1016/j.physrep.2007.01.001}

\bibitem[{{Mostert} {et~al.}(2021){Mostert}, {Duncan}, {R{\"o}ttgering},
  {Polsterer}, {Best}, {Brienza}, {Br{\"u}ggen}, {Hardcastle}, {Jurlin},
  {Mingo}, {Morganti}, {Shimwell}, {Smith}, \&
  {Williams}}]{2021A&A...645A..89M}
{Mostert}, R. I.~J., {Duncan}, K.~J., {R{\"o}ttgering}, H. J.~A., {et~al.}
  2021, \aap, 645, A89, \dodoi{10.1051/0004-6361/202038500}

\bibitem[{{Murante} {et~al.}(2015){Murante}, {Monaco}, {Borgani}, {Tornatore},
  {Dolag}, \& {Goz}}]{2015MNRAS.447..178M}
{Murante}, G., {Monaco}, P., {Borgani}, S., {et~al.} 2015, \mnras, 447, 178,
  \dodoi{10.1093/mnras/stu2400}

\bibitem[{{Murante} {et~al.}(2010){Murante}, {Monaco}, {Giovalli}, {Borgani},
  \& {Diaferio}}]{2010MNRAS.405.1491M}
{Murante}, G., {Monaco}, P., {Giovalli}, M., {Borgani}, S., \& {Diaferio}, A.
  2010, \mnras, 405, 1491, \dodoi{10.1111/j.1365-2966.2010.16567.x}

\bibitem[{{Norris} {et~al.}(2021{\natexlab{a}}){Norris}, {Intema},
  {Kapi{\'n}ska}, {Koribalski}, {Lenc}, {Rudnick}, {Alsaberi}, {Anderson},
  {Anderson}, {Crawford}, {Crocker}, {English}, {Filipovi{\'c}}, {Galvin},
  {Hopkins}, {Hurley-Walker}, {Inoue}, {Luken}, {Macgregor}, {Manojlovi{\'c}},
  {Marvil}, {O'Brien}, {Park}, {Raja}, {Shobhana}, {Venturi}, {Collier},
  {Hale}, {Hotan}, {Moss}, \& {Whiting}}]{2021PASA.38.3N}
{Norris}, R.~P., {Intema}, H.~T., {Kapi{\'n}ska}, A.~D., {et~al.}
  2021{\natexlab{a}}, \pasa, 38, e003, \dodoi{10.1017/pasa.2020.52}

\bibitem[{{Norris} {et~al.}(2021{\natexlab{b}}){Norris}, {Marvil}, {Collier},
  {Kapi{\'n}ska}, {O'Brien}, {Rudnick}, {Andernach}, {Asorey}, {Brown},
  {Br{\"u}ggen}, {Crawford}, {English}, {Rahman}, {Filipovi{\'c}}, {Gordon},
  {G{\"u}rkan}, {Hale}, {Hopkins}, {Huynh}, {HyeongHan}, {James Jee},
  {Koribalski}, {Lenc}, {Luken}, {Parkinson}, {Prandoni}, {Raja}, {Reiprich},
  {Riseley}, {Shabala}, {Sheil}, {Vernstrom}, {Whiting}, {Allison}, {Anderson},
  {Ball}, {Bell}, {Bunton}, {Galvin}, {Gupta}, {Hotan}, {Jacka}, {Macgregor},
  {Mahony}, {Maio}, {Moss}, {Pandey-Pommier}, \&
  {Voronkov}}]{2021PASA...38...46N}
{Norris}, R.~P., {Marvil}, J., {Collier}, J.~D., {et~al.} 2021{\natexlab{b}},
  \pasa, 38, e046, \dodoi{10.1017/pasa.2021.42}

\bibitem[{{Norris} {et~al.}(2022){Norris}, {Collier}, {Crocker}, {Heywood},
  {Macgregor}, {Rudnick}, {Shabala}, {Andernach}, {da Cunha}, {English},
  {Filipovi{\'c}}, {Koribalski}, {Luken}, {Robotham}, {Sekhar}, {Thorne}, \&
  {Vernstrom}}]{2022MNRAS.513.1300N}
{Norris}, R.~P., {Collier}, J.~D., {Crocker}, R.~M., {et~al.} 2022, \mnras,
  513, 1300, \dodoi{10.1093/mnras/stac701}

\bibitem[{{Omar}(2022{\natexlab{a}})}]{2022MNRAS.tmpL..74O}
{Omar}, A. 2022{\natexlab{a}}, \mnras, \dodoi{10.1093/mnrasl/slac081}

\bibitem[{{Omar}(2022{\natexlab{b}})}]{2022RNAAS...6..100O}
---. 2022{\natexlab{b}}, Research Notes of the American Astronomical Society,
  6, 100, \dodoi{10.3847/2515-5172/ac7044}

\bibitem[{{Ryu} {et~al.}(2019){Ryu}, {Kang}, \& {Ha}}]{2019ApJ...883...60R}
{Ryu}, D., {Kang}, H., \& {Ha}, J.-H. 2019, \apj, 883, 60,
  \dodoi{10.3847/1538-4357/ab3a3a}

\bibitem[{{Segal} {et~al.}(2022){Segal}, {Parkinson}, {Norris}, {Hopkins},
  {Andernach}, {Alexander}, {Carretti}, {Koribalski}, {Legodi}, {Leslie},
  {Luo}, {Pierce}, {Tang}, {Vardoulaki}, \& {Vernstrom}}]{2022arXiv220614677S}
{Segal}, G., {Parkinson}, D., {Norris}, R., {et~al.} 2022, arXiv e-prints,
  arXiv:2206.14677.
\newblock \doarXiv{2206.14677}

\bibitem[{{Sinha} \& {Holley-Bockelmann}(2009)}]{2009MNRAS.397..190S}
{Sinha}, M., \& {Holley-Bockelmann}, K. 2009, \mnras, 397, 190,
  \dodoi{10.1111/j.1365-2966.2009.14955.x}

\bibitem[{{Springel}(2005)}]{2005MNRAS.364.1105S}
{Springel}, V. 2005, \mnras, 364, 1105,
  \dodoi{10.1111/j.1365-2966.2005.09655.x}

\bibitem[{{Springel} {et~al.}(2001){Springel}, {White}, {Tormen}, \&
  {Kauffmann}}]{2001MNRAS.328..726S}
{Springel}, V., {White}, S. D.~M., {Tormen}, G., \& {Kauffmann}, G. 2001,
  \mnras, 328, 726, \dodoi{10.1046/j.1365-8711.2001.04912.x}

\bibitem[{{Teklu} {et~al.}(2015){Teklu}, {Remus}, {Dolag}, {Beck}, {Burkert},
  {Schmidt}, {Schulze}, \& {Steinborn}}]{2015ApJ...812...29T}
{Teklu}, A.~F., {Remus}, R.-S., {Dolag}, K., {et~al.} 2015, \apj, 812, 29,
  \dodoi{10.1088/0004-637X/812/1/29}

\bibitem[{{Valentini} {et~al.}(2017){Valentini}, {Murante}, {Borgani},
  {Monaco}, {Bressan}, \& {Beck}}]{2017MNRAS.470.3167V}
{Valentini}, M., {Murante}, G., {Borgani}, S., {et~al.} 2017, \mnras, 470,
  3167, \dodoi{10.1093/mnras/stx1352}

\bibitem[{{Valentini} {et~al.}(2020){Valentini}, {Murante}, {Borgani},
  {Granato}, {Monaco}, {Brighenti}, {Tornatore}, {Bressan}, \&
  {Lapi}}]{2020MNRAS.491.2779V}
---. 2020, \mnras, 491, 2779, \dodoi{10.1093/mnras/stz3131}

\bibitem[{{Vazza} {et~al.}(2012){Vazza}, {Br{\"u}ggen}, {van Weeren},
  {Bonafede}, {Dolag}, \& {Brunetti}}]{2012MNRAS.421.1868V}
{Vazza}, F., {Br{\"u}ggen}, M., {van Weeren}, R., {et~al.} 2012, \mnras, 421,
  1868, \dodoi{10.1111/j.1365-2966.2011.20160.x}

\bibitem[{{Wittor} {et~al.}(2021{\natexlab{a}}){Wittor}, {Ettori}, {Vazza},
  {Rajpurohit}, {Hoeft}, \&
  {Dom{\'\i}nguez-Fern{\'a}ndez}}]{2021MNRAS.506..396W}
{Wittor}, D., {Ettori}, S., {Vazza}, F., {et~al.} 2021{\natexlab{a}}, \mnras,
  506, 396, \dodoi{10.1093/mnras/stab1735}

\bibitem[{{Wittor} {et~al.}(2021{\natexlab{b}}){Wittor}, {Hoeft}, \&
  {Br{\"u}ggen}}]{2021arXiv211200023W}
{Wittor}, D., {Hoeft}, M., \& {Br{\"u}ggen}, M. 2021{\natexlab{b}}, arXiv
  e-prints, arXiv:2112.00023.
\newblock \doarXiv{2112.00023}

\bibitem[{{Zhang} {et~al.}(2020){Zhang}, {Churazov}, {Dolag}, {Forman}, \&
  {Zhuravleva}}]{2020MNRAS.494.4539Z}
{Zhang}, C., {Churazov}, E., {Dolag}, K., {Forman}, W.~R., \& {Zhuravleva}, I.
  2020, \mnras, 494, 4539, \dodoi{10.1093/mnras/staa1013}

\bibitem[{{Zou} {et~al.}(2019){Zou}, {Gao}, {Zhou}, \&
  {Kong}}]{2019ApJS..242....8Z}
{Zou}, H., {Gao}, J., {Zhou}, X., \& {Kong}, X. 2019, \apjs, 242, 8,
  \dodoi{10.3847/1538-4365/ab1847}

\end{thebibliography}
\bibliographystyle{aasjournal}

\end{document}